\newif\ifarxiv
\arxivtrue 

\ifarxiv
\documentclass[]{jfm-arxiv}
\else
\documentclass[lineno]{jfm}
\fi

\usepackage{graphicx}
\usepackage{newtxtext}
\usepackage{newtxmath}
\usepackage{natbib}
\usepackage{amsmath}
\usepackage{upgreek}
\usepackage{comment}
\usepackage{textcomp}
\usepackage{pifont}
\usepackage{hyperref}
\usepackage{booktabs}
\usepackage[dvipsnames]{xcolor}
\definecolor{myCyan}{rgb}{0.0, 1.0, 1.0}

\newcommand{\RomanNumeralCaps}[1]
\linenumbers
\newcommand{\comm}[1]{}
\newcommand{\q}[1]{\textquotedblleft #1\textquotedblright}
\shorttitle{Breaking of axisymmetry by an internal flow bifurcation}
\shortauthor{P. Shi, \'{E}. Climent, D. Legendre}

\title{Flow past a fixed spherical droplet: breaking of axisymmetry by an internal flow bifurcation}

\author{Pengyu Shi\aff{1,2} \corresp{\email{p.shi@hzdr.de}}, 
\'{E}ric Climent\aff{1} 
\and Dominique Legendre\aff{1} \corresp{\email{dominique.legendre@imft.fr}}}
	
\affiliation{\aff{1}Universit\'e de Toulouse; INPT, UPS; IMFT (Institut de M\'ecanique des Fluides de Toulouse), F-31400 Toulouse, France
\aff{2}Helmholtz-Zentrum Dresden – Rossendorf, Institute of Fluid Dynamics, 01328 Dresden, Germany}

\begin{document}
\maketitle

\begin{abstract}
Direct numerical simulations of a uniform flow past a fixed spherical droplet are performed to determine the parameter range within which the axisymmetric flow becomes unstable. The problem is governed by three dimensionless parameters: the drop-to-fluid dynamic viscosity ratio, $\mu^\ast$, and the external and internal Reynolds numbers, $\Rey^e$ and $\Rey^i$, which are defined using the kinematic viscosities of the external and internal fluids, respectively. 
The present study confirms the existence of a regime at low-to-moderate viscosity ratio where the axisymmetric flow breaks down due to an internal flow instability. In the initial stages of this bifurcation, the external flow remains axisymmetric, while the asymmetry is generated and grows only inside the droplet. As the disturbance propagates outward, the entire flow first transits to a biplanar symmetric flow, characterised by two pairs of counter-rotating streamwise vortices in the wake. A detailed examination of the flow field reveals that the vorticity on the internal side of the droplet interface is driving the flow instability. Specifically, the bifurcation sets in once the maximum internal vorticity exceeds a critical value that decreases with increasing $\Rey^i$. For sufficiently large $\Rey^i$, internal flow bifurcation may occur at viscosity ratios of $\mu^\ast = O(10)$, an order of magnitude higher than previously reported values. Finally, we demonstrate that the internal flow bifurcation in the configuration of a fixed droplet in a uniform fluid stream is closely related to the first path instability experienced by a buoyant, deformable droplet of low-to-moderate $\mu^\ast$ freely rising in a stagnant liquid.

\end{abstract}


\section{Introduction}
\label{sec:intro}
Single bubbles, droplets, and particles can experience complex paths when moving freely under the influence of gravity in an otherwise quiescent fluid \citep{2000_Magnaudet, 2005_Clift, 2012_Ern, mathai2020bubbly, 2024_Bonnefis}. Understanding the origin and nature of these irregular paths has been a longstanding concern in multiple disciplines, including mechanical or chemical engineering, aerodynamics, and meteorology. To a large extent, the onset of the first non-vertical path is closely related to the primary wake instability that occurs beyond a critical Reynolds number even if the body moves at a constant speed and orientation \citep{2012_Ern}. The first step in understanding path instability is to examine the conditions under which the axisymmetric wake of a fixed body with different boundary conditions (e.g., no-slip for particles and free-slip for bubbles) first becomes unstable \citep{dandy1989buoyancy, 1993_Natarajan, johnson1999flow, ghidersa2000breaking, yang2007linear, 2007_Magnaudet, tchoufag2013linear}. In this context, \cite{2007_Magnaudet} demonstrated that, regardless of the boundary conditions at the surface of the body, the axisymmetric wake becomes unstable when the maximum vorticity generated on the external side of the body surface exceeds a critical Reynolds-number-dependent threshold. Beyond this threshold, the steady wake exhibits a pair of counter-rotating trailing vortices generating a non-zero transverse lift force. Once the body is free to move, this force induces an oblique motion in the symmetry plane of the wake, ultimately leading to a non-vertical path \citep{mougin2002path, 2006_Mougin, 2010_Horowitz}.

The close connection between the primary wake instability behind a fixed body and the onset of the first non-vertical path when such a body is free to move has been well established for bubbles and particles, as well as droplets with a dynamic viscosity significantly higher than that of the surrounding fluid. For example, in the case of a high-Reynolds-number bubble whose surface is free of surfactants, the threshold for wake instability corresponds to a critical aspect ratio $\upchi\approx2.1$ (where $\upchi$ is the ratio of major and minor axes of the body) \citep{yang2007linear}, which is close to the threshold for path instability ($\approx2.0$) of a bubble that freely rises \citep{1995_Duineveld, 2008_Zenit, 2024_Bonnefis, 2025_Shi}. Similarly, for a solid sphere, where the no-slip boundary condition applies at the surface, the threshold for wake instability occurs at a critical Reynolds number of approximately 210, in good agreement with the range of the first path instability ($\in[210, 260]$) of a freely rising light sphere \citep{2004_Jenny, 2010_Horowitz, 2018_Auguste}. Lastly, for highly viscous droplets (i.e., those with $\mu^\ast\gg1$, where $\mu^\ast$ denotes the drop-to-fluid viscosity ratio) that behave similarly to solid particles, the relationship between wake and path instabilities appears evident. Specifically, \cite{2015_Albert} used direct numerical simulations (DNS) to investigate the path of corn oil droplets rising in pure water (for which $\mu^\ast\approx46$) and found that the path transition (from vertical to steady oblique) occurs at a critical Reynolds number equal approximately to 198, close to the primary wake instability threshold for a solid sphere ($\approx210$). The slightly lower critical Reynolds number observed is expected, as the droplet undergoes slight deformation ($\upchi \approx1.05$ at the threshold), leading to an increase of the external surface vorticity \citep{2007_Magnaudet} and, consequently, a reduction of the critical Reynolds number for wake instability.

The relationship between wake and path instabilities differs for droplets with low-to-moderate viscosity ratios [i.e., those with $\mu^\ast=\mathcal{O}(0.1-1)$]. For such droplets, the onset of the first path instability occurs at a significantly lower external surface vorticity (and thus a lower Reynolds number at a fixed aspect ratio, or \emph{vice versa}) than that predicted for the primary wake instability using the criterion proposed by \cite{2007_Magnaudet}. A representative example of this phenomenon is the experiment by \cite{wegener2009einfluss} [see also \cite{wegener2010terminal}], which investigated the motion of single Toluene droplets of various sizes rising in water. The corresponding drop-to-fluid viscosity ratio was $\mu^\ast = 0.62$ (see table 2 of \cite{wegener2010terminal} for detailed physical properties). For droplets with an equivalent radius exceeding $R\approx1.1\,\text{mm}$, the rising speed initially increased but then experienced a sudden decrease of approximately $30\%$, followed by pronounced oscillations around this reduced mean value. Furthermore, after several cycles of rising-speed oscillations, the path evolved from rectilinear to oblique \citep[see][Fig. 5.4 therein]{wegener2009einfluss}. Notably, at this threshold radius, the surface vorticity generated at the external side of the droplet in a fixed-droplet configuration (based on present DNS results, to be outlined in \S\,\ref{sec:mec_bif2}) is only one-third of that associated with the primary wake instability \citep[Eq.\,(4.1) therein]{2007_Magnaudet}. Similar observations for droplets with $\mu^\ast=\mathcal{O}(0.1-1)$—particularly the presence of a critical droplet size beyond which the (mean) terminal rising velocity undergoes a sudden reduction—have been reported in earlier experiments \citep{klee1956rate, thorsen1968terminal}. A comprehensive review of related studies can be found in \citet{abdelouahab2011new} and, more recently, in \cite{zhang2019state}.

Since the detailed experiments of \cite{wegener2009einfluss}, several attempts have been made using DNS to replicate the first path instability of Toluene droplets rising freely in water. Early studies in this direction \citep{2011_Baumler, engberg2014numerical, wegener2014numerical} carried out simulations in a two-dimensional axisymmetric configuration, assuming that the flow remains axisymmetric during the initial stage of the first path instability (i.e. before the path transitions to an oblique trajectory). In this constrained flow setup, the only possible cause of oscillations in the rising speed is the onset of axisymmetric deformations about the droplet's minor axis. In fact, once such shape oscillation modes become unstable, the rising speed oscillates as well, since both the vertical drag and the vertical added mass depend on the droplet cross-section and, consequently, its shape \citep{magnaudet2011reciprocal, lalanne2013effect, 2025_Shi}. However, predictions from these axisymmetric simulations indicated that the first unstable shape oscillation mode occurs beyond a critical size of $R\approx2.2\,\text{mm}$ \citep{2011_Baumler}, which is twice the size of the first path instability reported by \cite{wegener2009einfluss}. This discrepancy motivated subsequent studies to perform fully resolved three-dimensional simulations to better capture the first path instability \citep{bertakis2010validated, eiswirth2011experimental}. However, due to the slow development of axisymmetry-breaking processes (and thus the long physical time required), it was not possible until the recent works of \citet{2015_Engberg} and \citet{charin2019dynamic} that the first path instability at the threshold droplet size ($R\approx 1.1\,\text{mm}$) was reasonably replicated. Specifically, DNS results in the fully developed regime from \citet[Fig. 11 therein]{charin2019dynamic} revealed that for droplets with $R\geq 1\,\text{mm}$, two pairs of counter-rotating streamwise vortices form in the wake, as inferred from the three-dimensional velocity field at the rear of the droplet. This indicates that the axisymmetric wake has already broken down for $R\approx1\,\text{mm}$. However, it is important to note that beyond this threshold, the rising path may still remain rectilinear, since the transverse force remains zero because the wake, consisting of two pairs of vortex threads, retains its biplanar symmetry, an observation confirmed by the DNS results of \citet{charin2019dynamic}.

The \emph{biplanar} symmetric wake structure revealed by the DNS of \citet{charin2019dynamic} intuitively suggests that the instability responsible for the symmetry breaking of the flow around a droplet with a low-to-moderate viscosity ratio is associated with the azimuthal wavenumber $m=2$ \citep{ghidersa2000breaking, yang2007linear}. The mathematical nature of this symmetry breaking differs fundamentally from that observed in cases involving rising bubbles and settling or rising particles. In these cases, the first nonstraight path is typically triggered by a mode with azimuthal wavenumber $m=1$, leading to a transition from an axisymmetric to a \emph{uniplanar} symmetric flow state \citep{2004_Jenny, yang2007linear, tchoufag2013linear, 2024_Bonnefis}. Given this distinction, it is not surprising that the criterion for the primary wake instability established for the latter case by \cite{2007_Magnaudet} fails to predict the first path instability of Toluene droplets rising in water.

For a better understanding of the underlying physical mechanisms driving the first path instability of droplets with low-to-moderate viscosity ratios, it is first necessary to examine more systematically a simplified configuration — the wake instability of the flow past a fixed droplet over a wide range of viscosity ratio and Reynolds number. The first attempt in this direction appears to be the study by \citet{edelmann2017numerical}, which reported on three-dimensional simulations of uniform flow past a spherical droplet at Reynolds numbers of $\mathcal{O}(100)$. Interestingly, the authors reported that at a viscosity ratio of 0.5, the wake exhibited a biplanar symmetric structure, a feature that was later identified in the wake of freely rising droplets \citep{charin2019dynamic}. Subsequent and more systematic numerical investigations have been reported in the following papers \citep{rachih2019etude, gode2024flow, 2024_Shi_drop}. Specifically, \citet{gode2024flow} highlighted the significant role of the internal flow (i.e., the flow inside the droplet) in triggering the primary wake instability. In that work, the author observed an internal flow bifurcation for $\mu^\ast$ up to 2 under the constraint that the external flow remained axisymmetric. In fact, when the external flow was allowed to respond to this internal flow bifurcation, as shown in \citet{2024_Shi_drop}, the axisymmetric wake broke down, and the entire flow transitioned into a biplanar symmetric flow similar to that reported in \citet{edelmann2017numerical}. However, the critical viscosity ratio below which this internal bifurcation occurs is not fixed; rather, it varies within the range of 1 to 10 depending on the external and internal Reynolds numbers (definitions to be provided in the next section) \citep{2024_Shi_drop}.

Based on the previous studies commented above, two key issues remain open regarding the primary wake and path instabilities of droplets with low-to-moderate viscosity ratios:

\begin{itemize}
    \item \textbf{Fixed-droplet in a uniform flow:} The physical mechanisms driving the internal flow bifurcation \citep{gode2024flow, 2024_Shi_drop} remain unclear. Moreover, a criterion is still lacking for determining whether the axisymmetric flow is stable for a given set of parameters (in terms of Reynolds numbers and viscosity ratio). This criterion cannot be based solely on a fixed viscosity ratio, as demonstrated in \citet{2024_Shi_drop}.
    \item \textbf{Freely rising / settling case:} The direct relationship between the wake instability caused by the internal flow bifurcation in the fixed-droplet case and the first path instability of a freely rising / settling droplet has yet to be established. Specifically, for the well-documented first path instability of Toluene droplets freely rising in water, can we reasonably predict the critical droplet size using a criterion for the primary wake instability derived from studying fixed-droplet cases?
\end{itemize}

In this work, our aim is to address these two open issues, with a particular focus on the first—namely, the primary wake instability induced by internal flow bifurcation. To this end, we revisit the problem of flow instability past a spherical droplet and conduct direct numerical simulations over a wide range of dimensionless numbers using the JADIM code developed at IMFT \citep{legendre2019basset, rachih2019etude, rachih2020numerical, gode2023basset, gode2024flow}. 
The paper is organized as follows. In \S\,\ref{sec:problem_state}, we formulate the problem and outline the numerical approach. Section \ref{sec:overview} provides an overview of the results, highlighting the connection between the internal flow bifurcation and the vorticity generated on the internal side of the droplet surface. A typical transition sequence with increasing internal Reynolds number for fixed external Reynolds number and viscosity ratio is discussed in \S\,\ref{sec:tra_seq}. In \S\,\ref{sec:mec_bif}, we present a physical explanation for the mechanism driving the internal flow bifurcation. The relationship between the primary wake instability of a fixed droplet and the first path instability when the droplet is free to move is explored in \S\,\ref{sec:mec_bif2}, with a particular focus on the case of Toluene droplets rising in water. Based on the confirmed relationship, \S\, \ref{sec:reg_map} presents the threshold droplet size for the internal bifurcation of a nearly spherical droplet moving freely in water, using the criterion for internal bifurcation proposed in the present work. Conclusions, along with the perspectives arising from this study, are presented in \S\,\ref{sec:summ}.

\section{Problem statement and numerical approach}
\label{sec:problem_state}
We consider a spherical droplet of radius $R$, density $\rho^i$, and dynamic viscosity $\mu^i$ that is fixed in a Newtonian fluid of density $\rho^e$ and dynamic viscosity $\mu^e$. Far from the droplet interface, the external flow is a uniform stream along $\boldsymbol{e}_x$, described by $\boldsymbol{u}^\infty = u_{rel}\boldsymbol{e}_x$, where $u_{rel}$ represents the slip velocity of the external fluid relative to the droplet. The entire flow field is governed by the incompressible Navier–Stokes equations,
\begin{equation}
	\nabla\cdot \boldsymbol{u}^k=0, \quad
	\rho^k\left(\frac{\partial \boldsymbol{u}^k}{\partial t}+
	\boldsymbol{u}^k\cdot  \nabla\boldsymbol{u}^k\right)=
	- \nabla p^k +\nabla\cdot\boldsymbol{\uptau}^k,
	\label{eq:ns}
\end{equation}
where
$\boldsymbol{\uptau}^k=\mu^k\left( \nabla\boldsymbol{u}^k+{}^{T}\nabla\boldsymbol{u}^k \right)$ is the viscous part of the stress tensor $\boldsymbol{\Sigma}^k=-p^k\boldsymbol{I}+\boldsymbol{\uptau}^k$, and $\boldsymbol{u}^k$ and $p^k$ denote the disturbed velocity and pressure, respectively. Here, $k=i$ (likewise, $k=e$) refers to the fluid inside (outside) the droplet.

The boundary conditions are outlined below. At the surface of the droplet, the normal velocity must vanish due to the non-penetration condition, whereas the tangential velocity and shear stress must be continuous. These constraints yield the following boundary conditions at the droplet surface $r=R$:    
\refstepcounter{equation}
$$
\boldsymbol{u}^i\cdot \boldsymbol{n}=\boldsymbol{u}^e\cdot \boldsymbol{n}=0\,, \quad
\boldsymbol{n}\times \boldsymbol{u}^i=
      \boldsymbol{n}\times \boldsymbol{u}^e\,, \quad
\boldsymbol{n}\times (\boldsymbol{\uptau}^i\cdot \boldsymbol{n})=
      \boldsymbol{n}\times (\boldsymbol{\uptau}^e\cdot \boldsymbol{n}),
\eqno{(\theequation{a,b,c})}
\label{eq:BC1}
$$
where $r=(x^2+y^2+z^2)^{1/2}$ is the distance from the droplet centre, and $\boldsymbol{n}$ is the outward unit normal to the droplet surface. In the far field, we assume that the disturbance induced by the droplet vanishes, implies that $\boldsymbol{u}^k=\boldsymbol{u}^\infty$ as $r\to\infty$.

The steady-state solution of the problem is characterized by three dimensionless numbers: the viscosity ratio $\mu^\ast=\mu^i/\mu^e$, the external Reynolds number $\Rey^e$, and the internal Reynolds number $\Rey^i$. The latter two are defined as
\begin{equation}
\Rey^e=\frac{\rho^e u_{rel} (2R)}{\mu^e}, \quad
\Rey^i=\frac{\rho^i u_{rel} (2R)}{\mu^i}.
\label{eq:def_var}
\end{equation}
The drop-to-fluid density ratio can be expressed in terms of these three parameters as $\rho^\ast=\mu^\ast\,\Rey^i/\Rey^e$. In what follows, the viscosity ratio is varied from 0.01 to 100, allowing us to examine the evolution of the flow structure from the clean-bubble limit ($\mu^\ast \to 0$) to the solid-sphere limit ($\mu^\ast \to \infty$). At a given $\mu^\ast$, the two Reynolds numbers are varied independently, whereas in a real drop-liquid system, only one of them is independent once the drop-to-liquid density ratio $\rho^\ast$ is specified. Thus, arbitrarily varying $\Rey^i$ and $\Rey^e$ implies artificially changing the density ratio while keeping the viscosity ratio $\mu^\ast$ fixed (see the discussion section to relate this analysis to experiments).

The simulations were performed using the JADIM code developed at IMFT. This code has been previously applied to simulate the three-dimensional flow around spherical bubbles and particles, as well as the associated hydrodynamic forces \citep{legendre1998lift, adoua2009reversal, shi2020hydrodynamic, shi2021drag} and has been extended to compute three-dimensional flows around and inside spherical droplets \citep{legendre2019basset, rachih2020numerical, gode2023basset, 2024_Shi_drop}. The reader is referred to \citet{2024_Shi_drop} for details regarding the numerical implementation, including the mesh grid, boundary conditions, and validation tests confirming the reliability of the numerical approach. The only difference between the present study and that considered in \citet{2024_Shi_drop} is the type of undisturbed flow (the linear shear flow now being a uniform flow). It is worth noting that the mesh grid used in \citet{2024_Shi_drop}, which is also used in the present work, features highly refined grid cells near the droplet interface, with at least five nodes located within both the internal and external boundary layers for Reynolds numbers of up to 1000. This refinement ensures that both the internal and external flows are accurately resolved at high Reynolds numbers, particularly within the boundary layers on both sides of the droplet interface.

\section{Overview of the results} \label{sec:overview}
\subsection{Identification of internal and external flow bifurcations} \label{sec:id_bif}

The breaking of axisymmetry can be tracked by examining the perturbation energy, for which a convenient measure is the mean kinetic energy of the azimuthal velocity component (hereafter referred to as the azimuthal energy) \citep{thompson2001kinematics, 2007_Magnaudet}:

\begin{equation}
E^k=\frac{1}{\rho^e V_s u_{rel}^2} \int_{V^k}{\rho^k ||\boldsymbol{u}_\varphi^k||^2\mathrm{d}V^k},
\label{eq:energy}
\end{equation}
where $V_s=4 \pi R^3/3$ is the volume of the droplet, and $\boldsymbol{u}_\varphi^k$ is the azimuthal component of the local velocity. Here, $E^k$ (respectively $V^k$) with $k=i$ or $e$ denotes the azimuthal energy (respectively the domain) inside or outside the droplet. The azimuthal energy of the entire flow field is then given by $E= E^i +E^e$, which becomes positive as soon as the axisymmetry of the base flow breaks down.

\begin{figure}
\centerline{\includegraphics[scale=0.65]{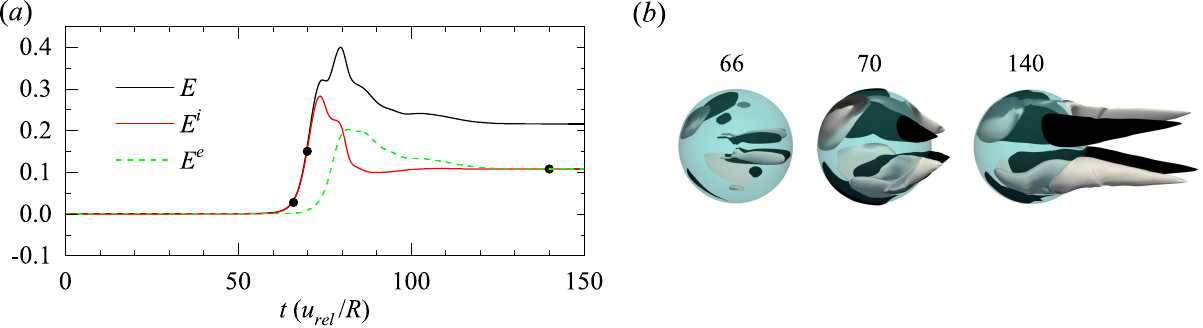}}
\caption{Characteristics of an internal flow bifurcation for $(\mu^\ast,\Rey^e,\Rey^i) = (0.5, 300, 1000)$. $(a)$ Total, internal, and external azimuthal energy as a function of time. $(b)$ Isosurfaces of the streamwise vorticity, $\omega_x R/u_{rel} = \pm 0.2$, at three selected time instants [indicated by numbers in $(b)$ and marked as bullets in $(a)$]. Grey and black threads correspond to positive and negative $\omega_x$, respectively.}
\label{fig:bif_int}
\end{figure}

Two different types of bifurcation can be identified based on the behaviour of $E^i$ and $E^e$. To illustrate this, we consider a series of cases with $(\Rey^e,\Rey^i) = (300, 1000)$ but with varying $\mu^\ast$. Figure \ref{fig:bif_int}$(a)$ presents the time evolution of the three azimuthal energy components for a low-viscosity-ratio droplet ($\mu^\ast = 0.5$). The axisymmetry of the base flow breaks down at $t\approx 60\,R/u_{rel}$, as indicated by the onset of growth in the total azimuthal energy $E$, which peaks at $t\approx 80\,R/u_{rel}$ before stabilising at a slightly lower value beyond $t\approx 140\,R/u_{rel}$. During this transition, $E^i$ maintains a substantial magnitude relative to $E$, while $E^e$ remains negligibly small in the initial stages (approximately for $t\,u_{rel}/R$ increasing from 60 to about 70). Figure \ref{fig:bif_int}$(b)$ displays isosurfaces of the streamwise vorticity $\omega_x$ at selected times during the transient. This vortical component becomes non-zero as soon as the bifurcation occurs. In the early stages of the bifurcation ($t=66\,R/u_{rel}$), $\omega_x$ is significant only inside the droplet, consistent with the initially negligible $E^e$ observed in panel $(a)$. As time progresses, the disturbance, represented by $\omega_x$ isocontours, grows and spreads outside the droplet at $t\approx 70\,R/u_{rel}$, leading to the formation of four vortex threads in the droplet wake. For the case under consideration, this wake structure remains stable in the fully 3D developed state.

\begin{figure}
\centerline{\includegraphics[scale=0.65]{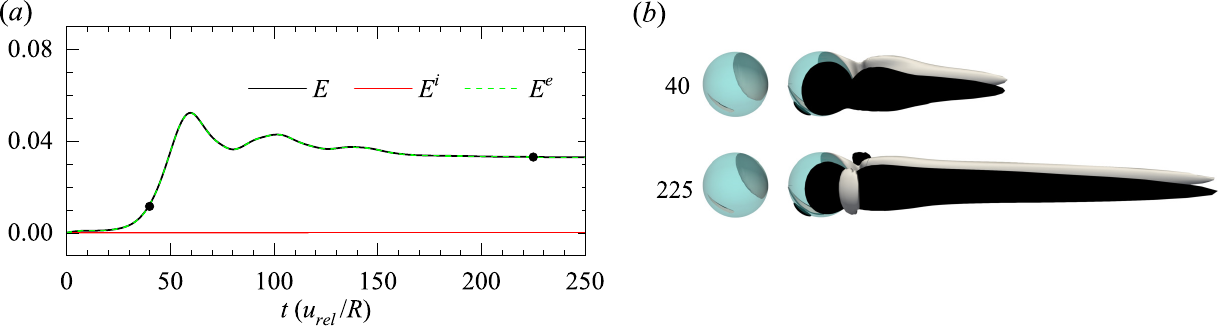}}
\caption{Same as figure \ref{fig:bif_int}, but for an external flow bifurcation in the case $(\mu^\ast,\Rey^e,\Rey^i) = (20, 300, 1000)$. In $(b)$, the isosurfaces correspond to $\omega_x R/u_{rel} = \pm 0.1$. The left part displays the vortical structure only inside the droplet and in the downstream half-space where the sign of $\omega_x$ in the wake is positive.}
\label{fig:bif_ext}
\end{figure}

For comparison, figure \ref{fig:bif_ext} presents the corresponding evolution for a droplet with a viscosity ratio of $\mu^\ast = 20$, which is 40 times larger than in the previous case. Compared with the low-$\mu^\ast$ case, the key difference in the evolution lies in the internal energy $E^i$, which now remains vanishingly small throughout the transition (figure \ref{fig:bif_ext}$a$). Additionally, the wake structure during the transition differs between the two cases. As shown in figure \ref{fig:bif_ext}$(b)$, the wake now comprises only one pair of streamwise vortices, in contrast to the two pairs observed in the low-$\mu^\ast$ case. From the perspective of linear dynamical systems theory \citep{ghidersa2000breaking, yang2007linear}, this wake structure indicates a symmetry breaking driven by a mode with azimuthal wavenumber $m=1$. In contrast, in the low-$\mu^\ast$ case, the presence of four vortex threads in the wake suggests that the symmetry breaking is caused instead by a mode with an azimuthal wavenumber $m=2$.

\begin{figure}
\centerline{\includegraphics[scale=0.595]{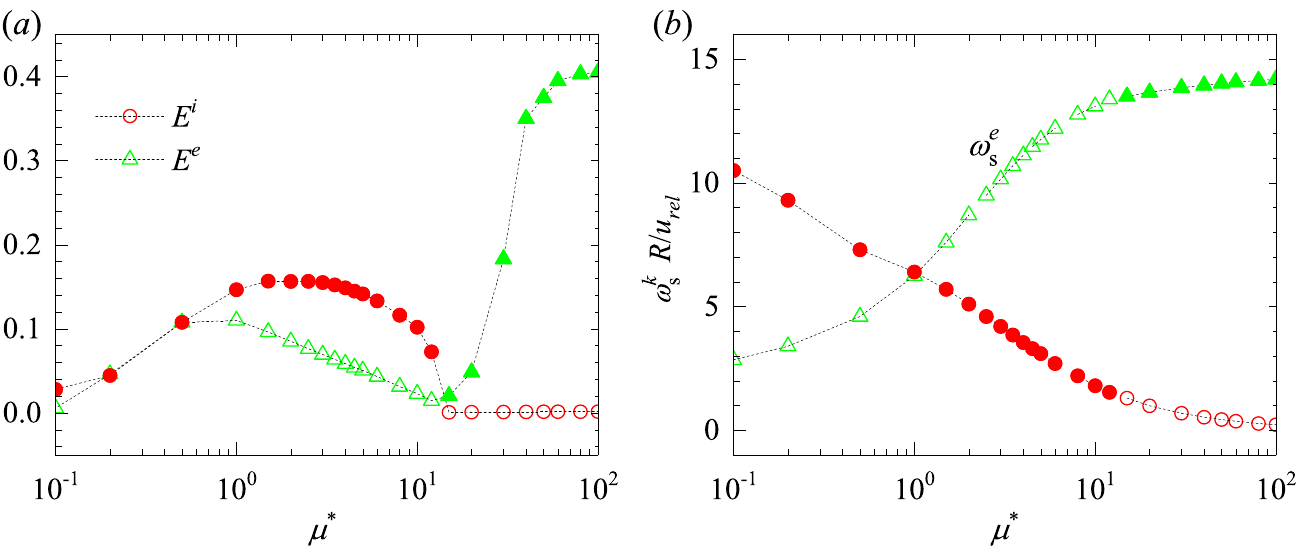}}
\caption{Results for $(a)$ the azimuthal energies $E^i$ and $E^e$ in the fully developed state and $(b)$ the maximum surface vorticity as a function of viscosity ratio $\mu^\ast$ obtained at steady state at $(\Rey^e,\Rey^i)= (300, 1000)$. In both panels, solid symbols in red (green) denote the onset of an internal (external) flow bifurcation.}
\label{fig:300e-1000i}
\end{figure}

Hereafter, we denote by an \emph{internal} bifurcation the type of bifurcation that occurs in the low-$\mu^\ast$ case, where the internal azimuthal energy $E^i$ remains significant throughout the transition. Conversely, we refer to as an \emph{external} bifurcation the instability occurring in the high-$\mu^\ast$ case, where $E^i$ remains vanishingly small throughout the transition. Figure \ref{fig:300e-1000i}$(a)$ summarises the results for $E^i$ and $E^e$ in the fully developed state obtained at $(\Rey^e,\Rey^i) = (300, 1000)$ over a wide range of $\mu^\ast$. Based on this classification, the internal bifurcation established for $\mu^\ast$ smaller than approximately 12, while the external bifurcation occurs for $\mu^\ast$ larger than about 15.

For a uniform flow past a solid particle or a clean bubble, the axisymmetry of the flow breaks down via the external bifurcation when the maximum vorticity generated on the external side of the body surface exceeds a critical $\Rey^e$-dependent value \citep{2007_Magnaudet}. To determine whether this empirical criterion also applies to a droplet, we examine the evolution of the maximum surface vorticity with the viscosity ratio. Since the surface vorticity is discontinuous at the interface (except when $\mu^\ast=1$), we introduce two distinct definitions of surface vorticity:
\begin{equation}
\boldsymbol\omega_S^i = \lim_{\;\: r \rightarrow R^-}\boldsymbol\omega^i - \left( \boldsymbol\omega^i \cdot \boldsymbol{n}\right) \boldsymbol{n}, \quad
\boldsymbol\omega_S^e = \lim_{\;\: r \rightarrow R^+}\boldsymbol\omega^e - \left( \boldsymbol\omega^e \cdot \boldsymbol{n}\right) \boldsymbol{n},
\label{eq:surf_vor}
\end{equation}
where $\boldsymbol\omega^k = \nabla \times \boldsymbol{u}^k$ is the vorticity, and $\boldsymbol\omega_S^i$ (respectively $\boldsymbol\omega_S^e$) denotes the surface vorticity on the internal (respectively external) side of the droplet interface. The maximum values of these quantities are denoted as $\omega_s^i$ and $\omega_s^e$, respectively, where $\omega_s^k = \max{\left(||\boldsymbol\omega_S^k||\right)}$. Unless stated otherwise, the results for $\omega_s^i$ and $\omega_s^e$ hereafter correspond to values obtained in the fully developed state of an imposed axisymmetric configuration, which are representative of the flow just prior to the onset of bifurcation.

Figure \ref{fig:300e-1000i}$(b)$ (green symbols) presents the maximum external surface vorticity $\omega_s^e$ as a function of $\mu^\ast$. Clearly, $\omega_s^e$ increases with increasing $\mu^\ast$ and exceeds approximately $13.5 u_{rel}/R$ (corresponding to the value at $\mu^\ast = 15$), beyond which the external bifurcation occurs. This threshold is close to the critical value ($13.8 u_{rel}/R$ for $\Rey^e = 300$) predicted by \citet[see Eq. (4.1) therein]{2007_Magnaudet}. Also shown in figure \ref{fig:300e-1000i}$(b)$ (red symbols) is the corresponding maximum internal surface vorticity $\omega_s^i$. In contrast to $\omega_s^e$, $\omega_s^i$ decreases with increasing $\mu^\ast$. The absence of an internal bifurcation for $\mu^\ast \gtrsim 12$ can be attributed to $\omega_s^i$ falling below a critical threshold. We will discuss this threshold in more detail in the next section.

\subsection{The internal flow bifurcation} \label{sec:io_bif}
In the previous section, we observed that for the series of cases with $(\Rey^e, \Rey^i) = (300, 1000)$, an internal bifurcation sets in when the viscosity ratio $\mu^\ast$ is smaller than approximately 12. However, the threshold $\mu^\ast$ can vary significantly with $\Rey^e$ and $\Rey^i$ \citep{edelmann2017numerical, gode2024flow, 2024_Shi_drop}, making it difficult to determine the internal bifurcation regime based solely on $\mu^\ast$. In this section, we demonstrate that the internal bifurcation is closely related to the maximum internal surface vorticity $\omega_s^i$ in the base flow and occurs when $\omega_s^i$ exceeds a critical value, which can be satisfactorily fitted using only the internal Reynolds number $\Rey^i$.
\begin{figure}
\centerline{\includegraphics[scale=0.57]{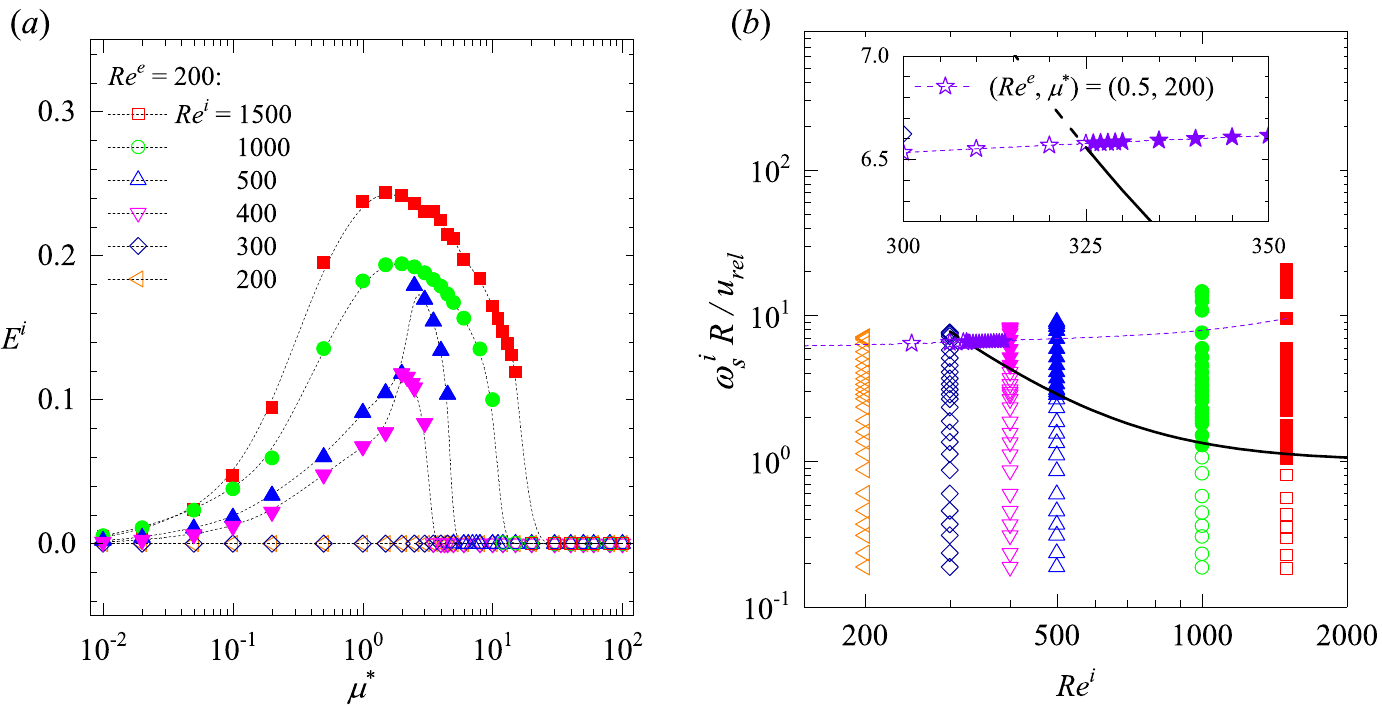}}
\caption{$(a)$ Internal azimuthal energy, $E^i$, in the fully developed state as a function of the viscosity ratio, $\mu^\ast$, for various $\Rey^i$ at $\Rey^e = 200$. $(b)$ Maximum internal surface vorticity as a function of $\Rey^i$. In $(b)$, in addition to the data at selected $\Rey^i$ values shown in $(a)$, an additional data series with increasing $\Rey^i$ for $(\mu^\ast, \Rey^e) = (0.5, 200)$ is also included (denoted by a thin dashed line and star symbols). In both panels, solid symbols indicate the onset of internal flow bifurcation. In $(b)$, for each iso-$\Rey^i$ data series, $\mu^\ast$ increases from top to bottom, and the thick black line represents the prediction from equation~\eqref{eq:cri_vor}.}
\label{fig:bif_re_200}
\end{figure}

We begin by examining the regime of internal bifurcation in the parameter space $(\mu^\ast, \Rey^i)$. To this end, we fix $\Rey^e$ at 200, ensuring that no external bifurcation occurs even in the solid-sphere limit $\mu^\ast \to \infty$ \citep{johnson1999flow, citro2016linear}. Figure \ref{fig:bif_re_200}$(a)$ presents the internal azimuthal energy $E^i$ in the fully developed state as a function of the viscosity ratio $\mu^\ast$ for various values of $\Rey^i$. Under the selected $\Rey^e$, no internal bifurcation occurs for $\Rey^i \leq 300$. However, at higher $\Rey^i$, the threshold $\mu^\ast$—below which the bifurcation sets in—increases from 3 at $\Rey^i = 400$ to 15 at $\Rey^i = 1500$. Figure \ref{fig:bif_re_200}$(b)$ shows the corresponding maximum internal surface vorticity, $\omega_s^i$. Unlike panel $(a)$, the results are plotted against $\Rey^i$ to highlight the dependence of the critical $\omega_s^i$ on $\Rey^i$. For each iso-$\Rey^i$ data series, $\omega_s^i$ decreases with increasing $\mu^\ast$, following a trend similar to that observed in figure \ref{fig:300e-1000i}$(b)$. These results indicate that the critical $\omega_s^i$ decreases as $\Rey^i$ increases. To evaluate the consistency of this trend, we carried out an additional series of simulations with increasing $\Rey^i$ while keeping $(\mu^\ast, \Rey^e)$ fixed at $(0.5, 200)$. The corresponding results for $\omega_s^i$ are shown in figure \ref{fig:bif_re_200}$(b)$ (dashed line and star symbols). Under this condition, the internal bifurcation takes place as $\Rey^i$ exceeds approximately 326, corresponding to a critical $\omega_s^i$ of about $6.56\,u_{rel}/R$.

\begin{figure}
\centerline{\includegraphics[scale=0.535]{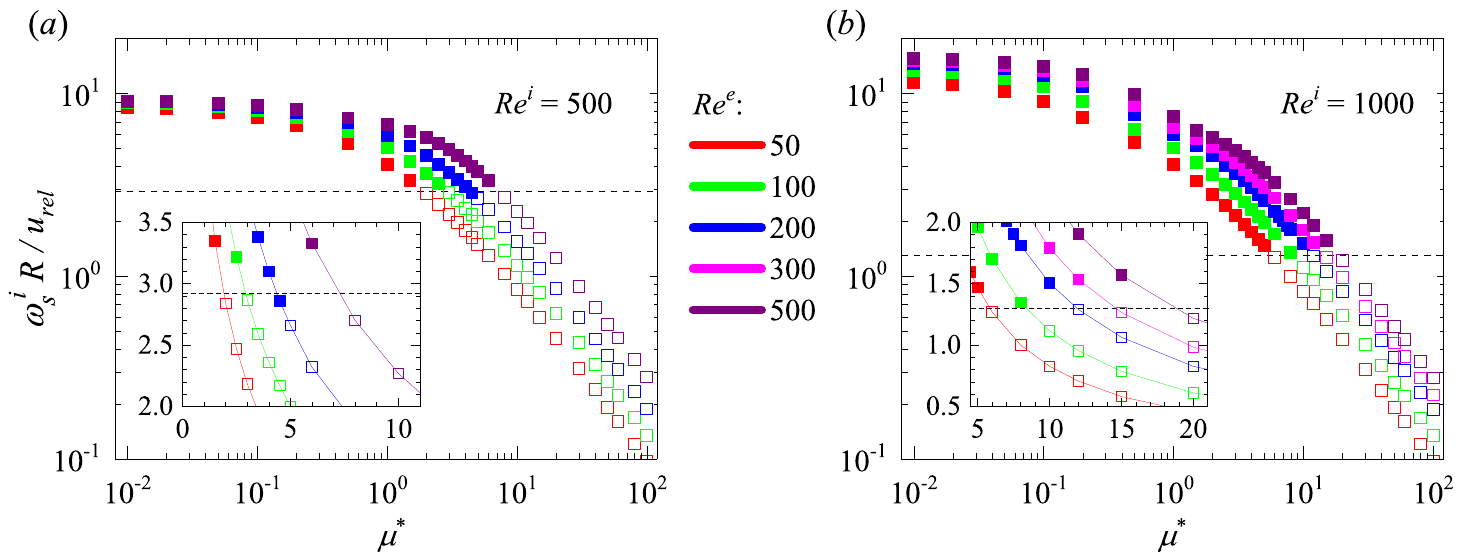}}
\caption{Maximum internal surface vorticity as a function of viscosity ratio $\mu^\ast$ for various $\Rey^e$ (distinguished by coloured symbols) at $(a)$ $\Rey^i=500$ and $(b)$ $\Rey^i=1000$. In both panels, solid symbols denote cases where internal bifurcation occurs, and the horizontal dashed line represents the corresponding $\omega_c^i(\Rey^i)$ according to equation \eqref{eq:cri_vor}.}
\label{fig:vor_i_500_1000}
\end{figure}

We collect the critical values of $\omega_s^i$, denoted as $\omega_c^i$, for the four data series with $\Rey^i > 300$ and fit them to a power-law relation in $\Rey^i$. The resulting empirical relation is
\begin{equation}
\omega_c^i\, R/u_{rel} \approx 1+0.33\left( \Rey^i/1000 \right)^{-2.5}.
\label{eq:cri_vor}
\end{equation}
Although equation \eqref{eq:cri_vor} is obtained for $\Rey^e = 200$, the predicted $\omega_c^i$ is generally applicable to different values of $\Rey^e$. To verify this, we formed two additional series of simulations: one at $\Rey^i=500$ and the other at $\Rey^i=1000$, varying $\Rey^e$ from 50 to 500 in both cases. Figure \ref{fig:vor_i_500_1000} presents the resulting $\omega_s^i$ as a function of the viscosity ratio, with cases involving the onset of internal bifurcation marked by solid symbols. According to equation \eqref{eq:cri_vor}, the critical $\omega_s^i$ for the internal bifurcation is approximately 2.9 at $\Rey^i=500$ and 1.3 at $\Rey^i=1000$. These two predictions are represented by horizontal dashed lines in figure \ref{fig:vor_i_500_1000}$(a,b)$, satisfactorily distinguishing the cases with internal bifurcation from those without.

Based on the discussion above, we may state that correlation \eqref{eq:cri_vor} can serve as an empirical criterion to determine whether the internal flow corresponding to a given set $(\mu^\ast, \Rey^e, \Rey^i)$ is stable or not. Specifically, given the maximum internal surface vorticity $\omega_s^i$ at the internal Reynolds number $\Rey^i$ under consideration, the internal flow is unstable (respectively, stable) if $\omega_s^i(\mu^\ast, \Rey^e, \Rey^i)$ is greater (respectively, smaller) than $\omega_c^i(\Rey^i)$. This correlation helps to explain the significant variation in the threshold viscosity ratio, $\mu_c^\ast$, for the internal bifurcation. In particular, since $\omega_c^i$ decreases with increasing $\Rey^i$, internal bifurcation is more likely to occur for droplets with larger $\Rey^i$. This explains why, in all previous studies \citep{edelmann2017numerical, rachih2019etude, gode2024flow, 2024_Shi_drop}, internal bifurcation has generally been observed at relatively large $\Rey^i$ (typically, $\Rey^i \geq 300$). On the other hand, since $\omega_c^i$ according to equation \eqref{eq:cri_vor} is independent of $\Rey^e$, while $\omega_s^i$ increases with $\Rey^e$, the threshold viscosity ratio $\mu_c^\ast$ at a given $\Rey^i$ increases with increasing $\Rey^e$, as shown in figure \ref{fig:vor_i_500_1000}.

\section{Transition sequence}\label{sec:tra_seq}

In this section, we focus on the series of cases with $(\mu^\ast, \Rey^e) = (0.5, 200)$ and examine how the flow structure evolves with increasing $\Rey^i$. This corresponds to varying the density ratio. In \S\ref{sec:mec_bif2}, we will show that the phenomena described below can indeed be observed for realistic physical properties of liquid–liquid systems. Since the whole problem depends on $(\mu^\ast, \Rey^e, \Rey^i)$, similar asymmetric flow structures can also arise with increasing $\Rey^e$ at a fixed $(\mu^\ast, \Rey^i)$ or with decreasing $\mu^\ast$ at fixed $(\Rey^e, \Rey^i)$. We did not explore in detail the bifurcation sequence under the latter two conditions, as doing so would require significant computational resources. However, it should be noted that while the critical $\Rey^e$ or $\mu^\ast$ for the first internal flow bifurcation can be reasonably predicted using the criterion \eqref{eq:cri_vor}, the sequence of higher-order bifurcations with increasing $\Rey^e$ or decreasing $\mu^\ast$ may differ from that observed with increasing $\Rey^i$.

\subsection{Axisymmetric flow regime} \label{sec:axi}
For the series of cases with $(\mu^\ast, \Rey^e) = (0.5, 200)$, our three-dimensional simulations indicate that the axisymmetry of the flow breaks down through an internal bifurcation as $\Rey^i$ exceeds a critical value approximately equal to 326 (see the star symbols in the inset of figure \ref{fig:bif_re_200}$b$).
\begin{figure}
\centerline{\includegraphics[scale=0.65]{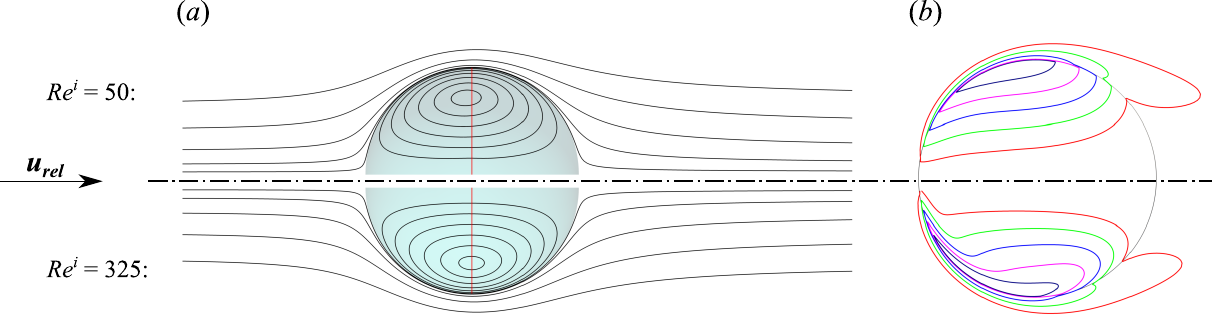}}
\caption{$(a)$ Streamlines and $(b)$ isocontours of the azimuthal vorticity $\omega_\phi$ around the droplet for $\Rey^i = 50$ (top panels) and $\Rey^i = 325$ (bottom panels). For both cases, $(\mu^\ast, \Rey^e) = (0.5, 200)$. In $(a)$, the vertical red line denotes $x = 0$. In $(b)$, coloured lines represent $-\omega_\phi R/u_{rel} = 1$ (red), 2 (green), 3 (blue), 4 (magenta), and 5 (navy).}
\label{fig:axi_streamline}
\end{figure}
Figure \ref{fig:axi_streamline}$(a)$ illustrates the streamlines around the droplet for $\Rey^i = 50$ (top panel) and $\Rey^i = 325$ (bottom panel), with the latter corresponding to the case just before the onset of bifurcation. In both cases, the external streamlines at the rear remain attached to the droplet, indicating the absence of a standing eddy in the wake prior to the onset of internal bifurcation. The internal flow structure resembles a Hill (spherical) vortex \citep{hill1894vi}, although a fore-aft asymmetry with respect to $x = 0$ (marked by the red vertical line in figure \ref{fig:axi_streamline}$a$) can be inferred from the results at $\Rey^i = 50$. This fore-aft asymmetry becomes more evident when examining the isocontours of the azimuthal vorticity $\omega_\phi$, as shown in figure \ref{fig:axi_streamline}$(b)$. Near the front of the droplet, the internal isocontours tilt toward the stagnation point, instead of aligning horizontally along the symmetry axis as in a Hill vortex. This tilting is more pronounced at higher $\Rey^i$: close to the stagnation point, the edges of the internal isocontours align almost parallel to the droplet surface. This strong tilting of azimuthal vorticity has also been observed in the wake of an oblate spheroidal bubble \citep{2007_Magnaudet, yang2007linear}, where the external $\omega_\phi$-isocontours tilt so that they align nearly perpendicular to the symmetry axis at the threshold of the external bifurcation.

\subsection{Biplanar-symmetric flow regime} \label{sec:bi_regime}

\begin{figure}
\centerline{\includegraphics[scale=0.57]{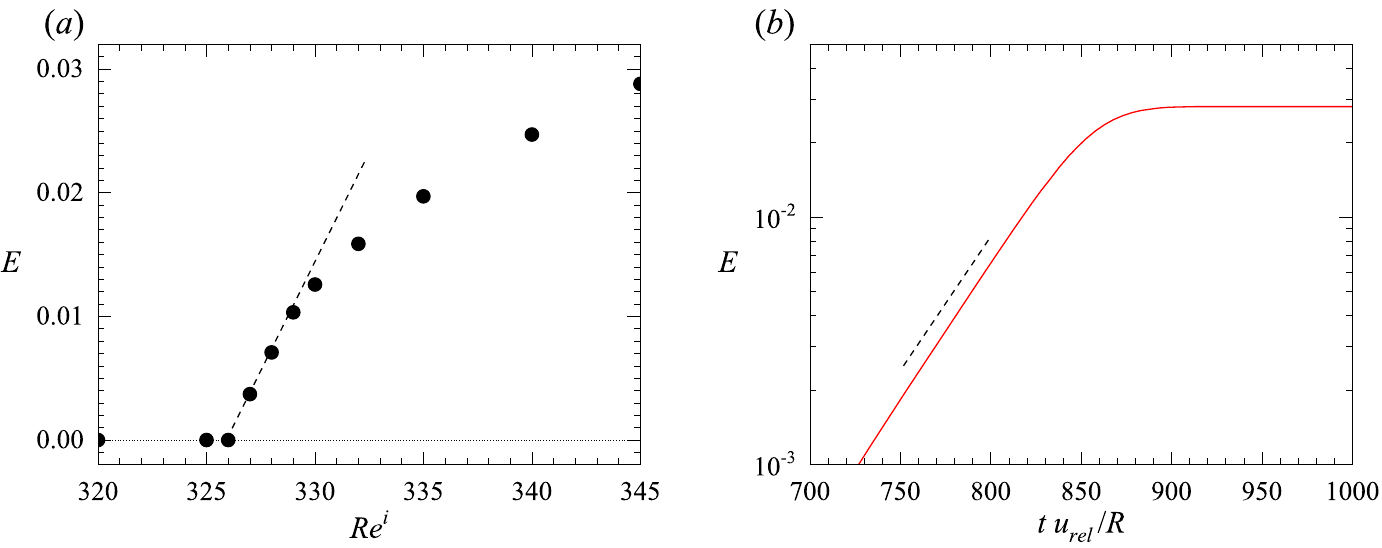}}
\caption{$(a)$ Variation of the total azimuthal energy of the steady state, $E$, with the internal Reynolds number $\Rey^i$ close to the threshold. $(b)$ $E$ as a function of time for $\Rey^i = 345$. In both panels, the straight dashed line highlights the linear scaling.}
\label{fig:e_tot_bip}
\end{figure}

The internal flow bifurcation sets in beyond $\Rey^i \approx 326$ for $(\mu^\ast, \Rey^e) = (0.5, 200)$. Following this bifurcation, the flow transits from an axisymmetric to a biplanar-symmetric structure (figure \ref{fig:bif_int}$b$). Once the axisymmetry breaking has saturated, the flow in all cases within this regime remains steady, indicating that the bifurcation is regular. Figure \ref{fig:e_tot_bip}$(a)$ presents the total azimuthal energy in the final state as a function of $\Rey^i$. The results clearly indicate that the bifurcation is supercritical. Figure \ref{fig:e_tot_bip}$(b)$ shows the time evolution of the total azimuthal energy at $\Rey^i = 345$. After linear transient growth (marked with a dashed straight line), the initial deviation from linearity levels off with a decreasing growth rate, further confirming the supercritical nature of the bifurcation \citep[p. 82]{strogatz2018nonlinear}.

\begin{figure}
\centerline{\includegraphics[scale=0.65]{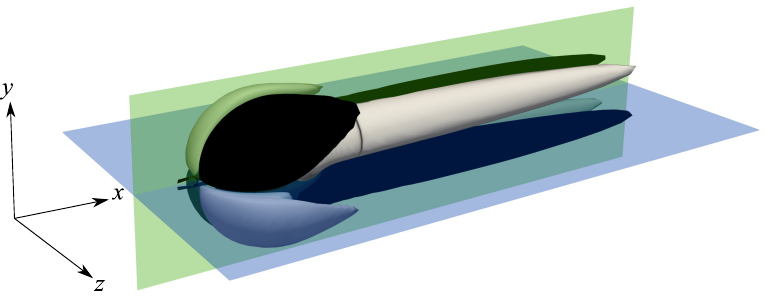}}
\caption{Isosurfaces of the streamwise vorticity, $\omega_x R/u_{rel} = \pm0.05$, past a droplet at $\Rey^i=345$ (grey and black threads correspond to positive and negative values, respectively). The flat surface in green (respectively, blue) highlights the symmetry plane in which the flow diverges (respectively, converges).}
\label{fig:345i_vor_x}
\end{figure}

Figure \ref{fig:345i_vor_x} presents the streamwise vorticity structure in the fully developed state for $\Rey^i = 345$. The resulting configuration, consisting of four vortex threads of equal intensity, closely resembles that observed in figure \ref{fig:bif_int}$(b)$. Due to the entrainment of fluid elements by these vortex threads, two distinct symmetry planes exist. The first, denoted as the $y=0$ plane (coloured blue), is characterised by the inward motion of fluid elements towards the symmetry axis. The second plane, denoted as the plane $z=0$ (coloured green), is associated with an outward motion away from the symmetry axis. Note that the positions of the two symmetry planes are determined by the initial disturbance. In our numerical setup, a weak streamwise linear shear flow of $10^{-4}\, y (u_{rel}/R)\, \boldsymbol{e}_x$ is imposed to trigger the bifurcation, thereby prescribing the locations of the two symmetry planes. The velocity variation across the droplet scale is only $10^{-4}u_{rel}$, making it negligibly small compared to the ambient flow.  

\begin{figure}
\centerline{\includegraphics[scale=0.65]{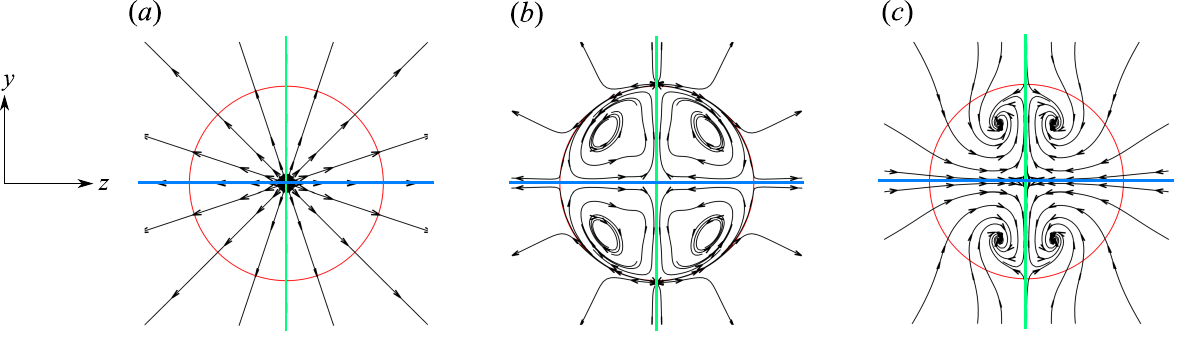}}
\caption{Two-dimensional streamlines of the disturbance $\boldsymbol{u}^k - u_x^k \boldsymbol{e_x}$ in selected cross-stream planes for $\Rey^i=345$. $(a)$, $(b)$, and $(c)$ correspond to $x/R = -2$, 0, and 5, respectively. In each panel, the red circle represents the boundary of a cylindrical surface $(y^2+z^2)^{1/2} = R$. The thick horizontal blue line (respectively, vertical green line) denotes the symmetry plane $y=0$ (respectively, $z=0$), as shown in figure \ref{fig:345i_vor_x}.}
\label{fig:345i_streamline_x}
\end{figure}

To further illustrate the biplanar symmetry of the flow structure, figure \ref{fig:345i_streamline_x} presents the two-dimensional streamlines of the disturbance in selected cross-stream planes. The disturbance is obtained by subtracting the streamwise velocity component from the full velocity field, i.e., the plotted streamlines correspond to $\boldsymbol{u}^k - u_x^k \boldsymbol{e_x}$. At a distance of $2R$ upstream of the droplet (figure \ref{fig:345i_streamline_x}$a$), all streamlines radiate outward from the symmetry axis connected to the front stagnation point, indicating that the flow remains axisymmetric at this location. At the cross-stream plane passing through the droplet centre (panel $b$), the axisymmetry of the inside base flow effectively breaks into four energetic vortices. The internal flow structure closely resembles that associated with the azimuthal wavenumber $m=2$ in the wake of a solid sphere \citep[see Fig. 13$(c)$ in][]{ghidersa2000breaking}. As this asymmetric internal disturbance influences downstream flows (panel $c$), vortex pairs at the same $y$-level gradually deviate from the symmetry plane $y=0$ while simultaneously converging towards each other.

\subsection{Uniplanar-symmetric flow regime} \label{sec:uni_regime}

A secondary bifurcation occurs as $\Rey^i$ exceeds $\approx 370$ for $(\mu^\ast, \Rey^e) = (0.5, 200)$. This new bifurcation disrupts the biplanar-symmetric flow structure, resulting in a flow with a single plane of symmetry in the fully developed state.

\begin{figure}
\centerline{\includegraphics[scale=0.575]{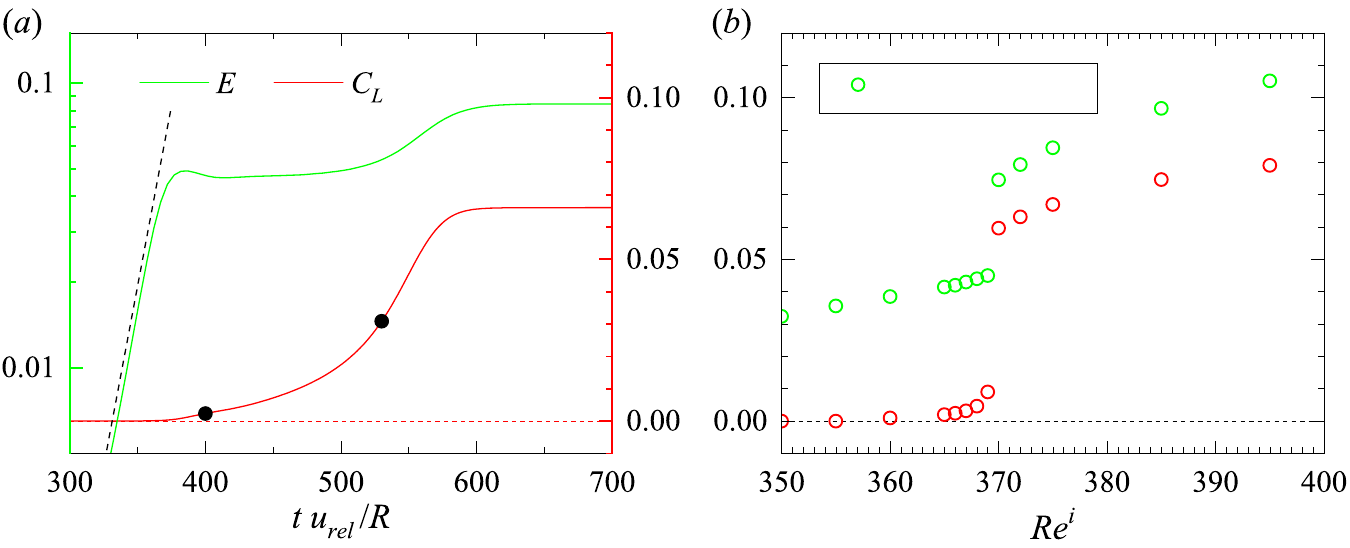}}
\caption{$(a)$ The azimuthal energy, $E$, and the lift coefficient, $C_L$, as functions of time for $\Rey^i = 375$. $(b)$ Variation of $E$ and $C_L$ in the fully developed state with the internal Reynolds number $\Rey^i$ for $\Rey^i$ ranging from 350 to 400. In $(a)$, the dashed black line highlights the linear growth rate of $E$, and the two bullets indicate the instants at $t\,u_{rel}/R = 400$ and 530, respectively. }
\label{fig:375i_e_cl}
\end{figure}

\begin{figure}
\centerline{\includegraphics[scale=0.65]{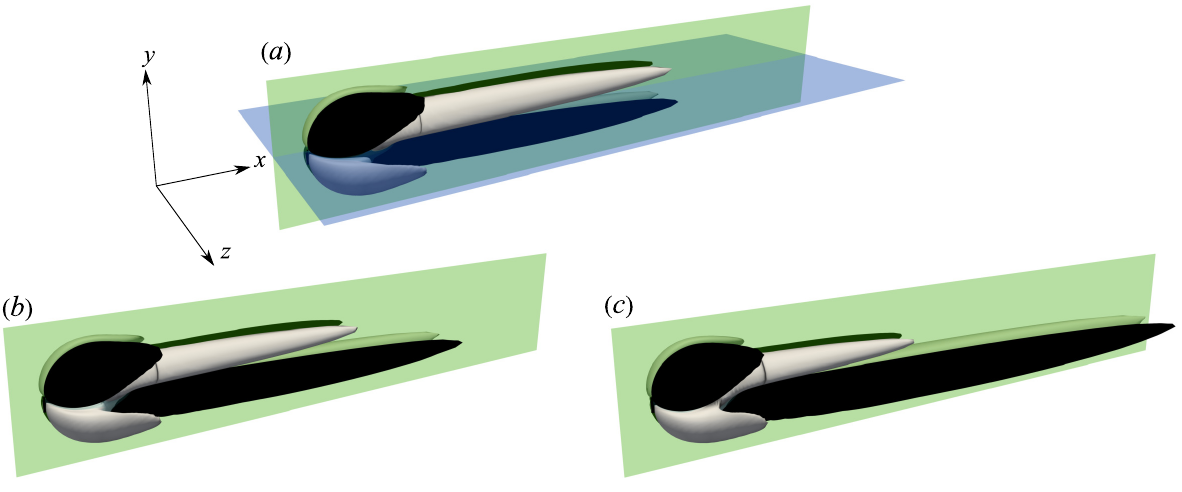}}
\caption{Isosurfaces of the streamwise vorticity, $\omega_x R/u_{rel} = \pm0.05$, past a droplet at selected time instants for $\Rey^i=375$. $(a)$, $(b)$, and $(c)$ correspond to $t\,u_{rel}/R = 400$, 530, and 2000, respectively.}
\label{fig:375i_vor_x}
\end{figure}

Taking the case at $\Rey^i=375$ as an example, figure \ref{fig:375i_e_cl}$(a)$ shows the time evolution of the total energy $E$ during this transition. The primary bifurcation sets in at $t\,u_{rel}/R \approx 300$, from which $E$ initially exhibits linear growth before decreasing and temporarily reaching a first plateau value as $t\,u_{rel}/R$ exceeds $\approx 400$. Figure \ref{fig:375i_vor_x}$(a)$ illustrates the vortical structure at this time. The flow remains biplanar-symmetric, similar to that observed in the fully developed state for $\Rey^i = 345$ (see figure \ref{fig:345i_vor_x}). Until this time, the evolution of energy and flow structure closely resembles that of the biplanar flow regime, confirming that the primary bifurcation remains supercritical. However, at $\Rey^i=375$, the biplanar-symmetric flow structure is unstable. As shown in figure \ref{fig:375i_e_cl}$(a)$, shortly after stabilizing at the first plateau, $E$ again increases and when $t\,u_{rel}/R$ exceeds 650 it re-stabilizes at a second plateau value approximately twice as high as the first one. Figure \ref{fig:375i_vor_x}$(b,c)$ depicts the vortical structures at two instants: one during the secondary transition ($t\,u_{rel}/R=530$; panel $b$) and the other in the fully developed state ($t\,u_{rel}/R=2000$; panel $c$). These results reveal that during the transition, the flow symmetry with respect to $y=0$ breaks down. Specifically, one of the vortex pairs at the same $y$ level (the pair with $y>0$ in figure \ref{fig:375i_vor_x}) shrinks while the other grows over time. Throughout this evolution, the flow symmetry with respect to $z=0$ persists.

The flow asymmetry with respect to $y=0$ during the second transition results in a non-zero lift force $F_L$ along the $y$-axis, directed from the primary vortex pair towards the shrinking one. We define this direction as the positive $y$-axis. The resulting lift force can be quantified using a lift coefficient, defined as $F_L = C_L \pi R^2 \rho^e u_{rel}^2 / 2$. Figure \ref{fig:375i_e_cl}$(a)$ (red line) shows the time evolution of the lift coefficient. The growth of the lift force begins at $t\,u_{rel}/R \approx 375$, well before $E$ reaches its first plateau. This indicates that the secondary bifurcation sets in before the primary bifurcation fully saturates. However, the initial growth rate of $C_L$ remains small. Beyond $t\,u_{rel}/R \approx 450$, $C_L$ starts to increase progressively, reaching approximately 0.065 in the fully developed state. Note that this value is close to that induced by the external flow bifurcation in the case of a solid sphere moving at $\Rey^e = 250$ \citep{johnson1999flow, shi2021drag}. Figure \ref{fig:375i_e_cl}$(b)$ shows the second plateau values of the total energy and the lift coefficient for $\Rey^i$ up to 400. Notably, both parameters exhibit a sharp increase as $\Rey^i$ exceeds approximately 370, suggesting that, unlike the primary bifurcation, the secondary bifurcation is not supercritical. This aspect will be discussed further in the next section, where we examine the stability of the dynamical system under finite-amplitude initial disturbances.

\subsection{Bistable flow regime} \label{sec:bistab_regime}
\begin{figure}
\centerline{\includegraphics[scale=0.55]{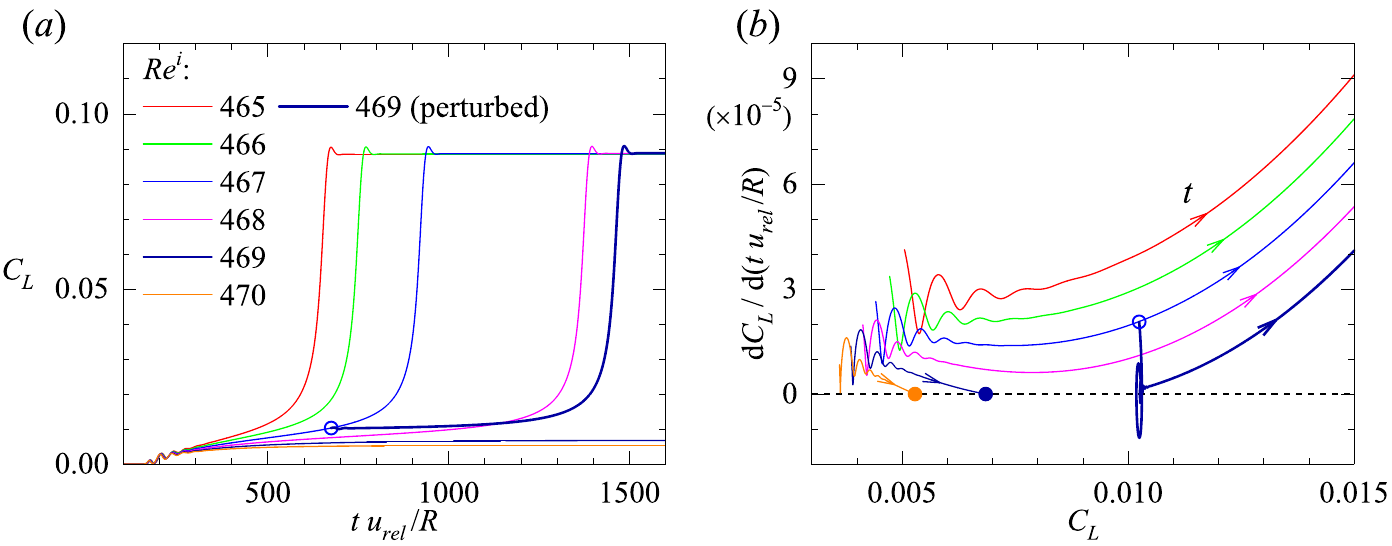}}
\caption{$(a)$ Time evolution of the lift coefficient for internal Reynolds numbers near the transition to the bistable regime. $(b)$ Variation of $\mathrm{d}C_L(t)/\mathrm{d}\left(t u_{rel}/R \right)$ as a function of $C_L(t)$ over the time interval where $C_L(t)$ evolves slowly (results for $t u_{rel}/R \leq 300$ are omitted). For all considered $\Rey^i$, the simulation starts from an initially unperturbed state. For $\Rey^i=469$, an additional simulation (labelled as \q{perturbed}) was performed, starting from an initially asymmetric state based on a result from $\Rey^i=467$ at $t u_{rel}/R = 675$ (denoted by a blue symbol in both panels). In $(b)$, the two cases with $\Rey^i > 468$ approach stable fixed points with small but finite $C_L$ (denoted by solid symbols), whereas in the remaining cases, $C_L$ after escaping the bottleneck continues to increase with time (as indicated by the arrows).}
\label{fig:cl_rate_bistable}
\end{figure}

As $\Rey^i$ increases further while keeping $(\mu^\ast, \Rey^e) = (0.5, 200)$, the primary bifurcation occurs earlier in time. For instance, at $\Rey^i = 450$, the azimuthal energy $E$ starts to increase at approximately $150\,R/u_{rel}$ (not shown), compared with about $300\,R/u_{rel}$ for $\Rey^i = 375$ (see figure \ref{fig:375i_e_cl}$a$). In contrast, the secondary bifurcation becomes progressively slower and eventually ceases to occur for $\Rey^i > 468$. This latter behaviour is highlighted in figure \ref{fig:cl_rate_bistable}$(a)$, which shows the time evolution of $C_L$ for cases near this transition. For $\Rey^i \leq 468$, $C_L$ initially rises slowly to a plateau before gradually increasing to its final value. However, for larger $\Rey^i$ cases, $C_L$ settles at a plateau that decreases with increasing $\Rey^i$ and remains at this level as $t\,u_{rel}/R \to \infty$. The $C_L(t)$ evolutions resemble systems close to a saddle-node bottleneck \citep[Sec. 4.3] {strogatz2018nonlinear}, where a saddle-node remnant or \emph{ghost} induces slow passage. To better illustrate this, figure \ref{fig:cl_rate_bistable}$(b)$ presents the rate of change of the lift coefficient, $\mathrm{d}C_L(t)/\mathrm{d}\left(t u_{rel}/R \right)$, as a function of $C_L(t)$ during the interval where $C_L$ varies slowly over time. A bottleneck causing slow passage is clearly visible at $\Rey^i=466$, shrinking as $\Rey^i$ increases. For $\Rey^i > 468$, the \q{path} originating from $C_L=0$ terminates at a \emph{stable} fixed point with a small but finite lift coefficient (indicated by solid symbols in panel $b$).

\begin{figure}
\centerline{\includegraphics[scale=0.65]{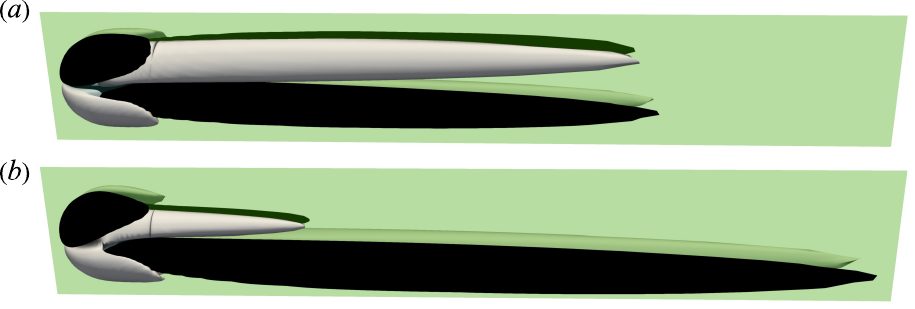}}
\caption{Isosurfaces of the streamwise vorticity, $\omega_x R/u_{rel} = \pm0.05$, in the fully developed state for $\Rey^i=469$ corresponding to different initial conditions. $(a)$ Simulation starting from an initially axisymmetric state. $(b)$ Simulation starting from a slightly asymmetric state derived from the transient result for $\Rey^i=467$, where $C_L=0.0102$ (denoted by an open circle in figure \ref{fig:cl_rate_bistable}).}
\label{fig:469i_vor_x}
\end{figure}

\begin{figure}
\centerline{\includegraphics[scale=0.65]{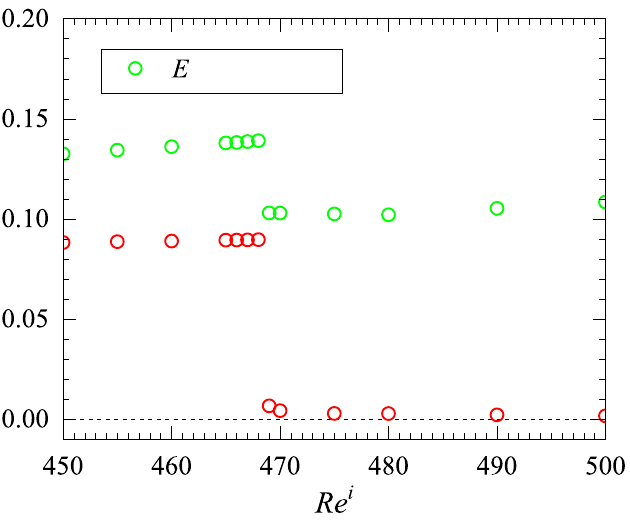}}
\caption{Variation of azimuthal energy $E$ and lift coefficient $C_L$ in the fully developed state for $\Rey^i$ increasing from 450 to 500. All Simulations started with an initially axisymmetric flow.}
\label{fig:cl_e_bistable1}
\end{figure}

Figure \ref{fig:469i_vor_x} illustrates the streamwise vorticity structure in the fully developed state for $\Rey^i=469$. Given the small $C_L$ in this case, the flow structure remains nearly biplanar symmetric, albeit with a slight up-down asymmetry. Figure \ref{fig:cl_e_bistable1} presents the azimuthal energy and lift coefficient in the fully developed state for $\Rey^i$ increasing from 450 to 500. Due to the bottleneck effects mentioned above, both $E$ and $C_L$ undergo an abrupt decrease as $\Rey^i$ exceeds approximately 468.

We are now in a position to examine the response of the flow in the presence of a finite-amplitude initial disturbance. This is necessary because, as mentioned in the last part of \S\,\ref{sec:uni_regime}, the secondary bifurcation is not supercritical, and multiple stable states may coexist depending on the initial conditions. To begin with, we revisit the cases near $\Rey^i = 468$. All cases were initialized from an axisymmetric state where $C_L = 0$. To investigate the possible existence of an additional stable state, we performed the simulation with $\Rey^i=469$ again starting from a slightly asymmetric initial state with $C_L=0.0102$. This \q{initial} state was derived from the transient result for $\Rey^i=467$ at the instant when the system had just passed the bottleneck (denoted by an open symbol in figure \ref{fig:cl_rate_bistable}$b$). As seen in figure \ref{fig:cl_rate_bistable}$(b)$ (thick line), even with such a small initial disturbance, the system was able to \q{escape} the bottleneck. Following this escape, $C_L$ increased sharply to its final level, nearly identical to the values reached at slightly smaller $\Rey^i$ (thick line in figure \ref{fig:cl_rate_bistable}$a$). Figure \ref{fig:469i_vor_x}$(b)$ shows the structure of the streamwise vorticity in the fully developed state. The flow exhibits a strong up-down asymmetry, significantly different from the flow structure obtained when starting from an initially axisymmetric state (panel $a$), but similar to that shown in figure \ref{fig:375i_vor_x}$(c)$.

The test above indicates that for $\Rey^i > 468$, the flow may evolve toward at least two distinct asymmetric branches. One, in which the flow remains nearly biplanar-symmetric, appears to be stable only to small disturbances. The other, in which the flow exhibits a uniplanar-symmetric structure, may be triggered by larger amplitude of \q{up-down} disturbances (hence $C_L > 0$). Hereafter, these two asymmetric branches will be referred to as the biplanar and uniplanar branches, respectively. So far, all cases discussed (except one) started from an initially axisymmetric state, corresponding to a vanishingly small initial disturbance. To examine the response of the flow system to \q{large} disturbances, we carried out  these cases again starting from a uniplanar-symmetric state reached in the fully developed stage of $\Rey^i=450$. For this new \q{initial} state, the up-down asymmetry of the flow leads to a lift coefficient equal to $C_L = 0.088$, providing an initial disturbance that we believe is large enough to trigger the transition to the uniplanar branch. 
\begin{figure}
\centerline{\includegraphics[scale=0.595]{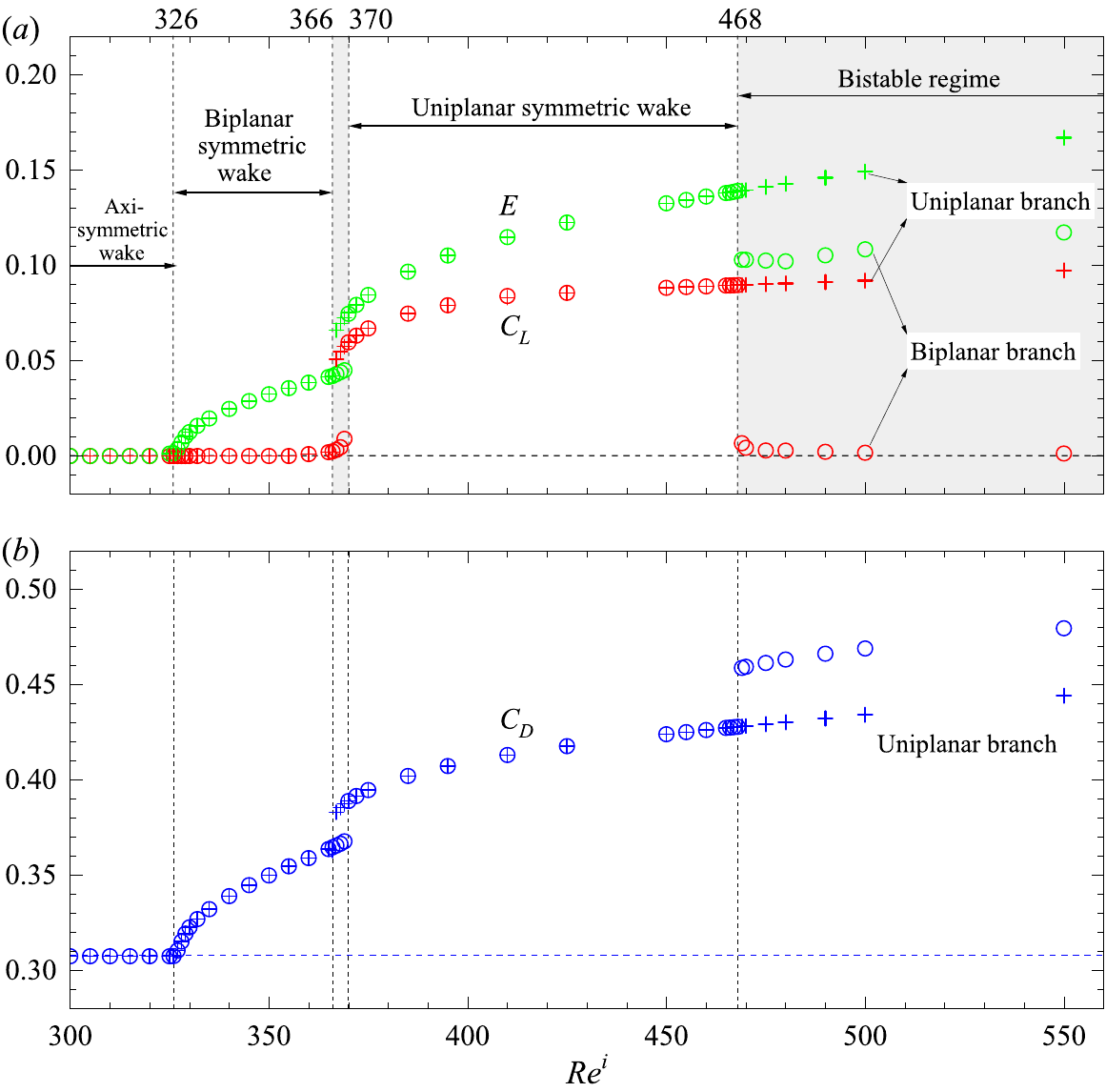}}
\caption{Azimuthal energy $E$ (green symbols), lift coefficient $C_L$ (red symbols), and drag coefficient $C_D$ (blue symbols) in the fully developed state for $\Rey^i$ increasing from 300 to 550 with $(\mu^\ast, \Rey^e) = (0.5, 200)$. $\bigcirc$: simulations starting with an initially axisymmetric flow; $+$: simulations starting with a uniplanar-symmetric flow corresponding to the fully developed state at $\Rey^i = 450$. Vertical dashed lines highlight the critical $\Rey^i$ values marking regime transitions. In $(a)$, the two shaded grey regions correspond to the two bistable regimes. In $(b)$, the horizontal dashed line denotes the drag coefficient obtained by enforcing axisymmetry regardless of $\Rey^i$.}
\label{fig:cl_e_bistable2}
\end{figure}

Figure \ref{fig:cl_e_bistable2}$(a)$ (cross symbols) presents the total energy and lift coefficient in the fully developed state for $\Rey^i$ increasing from 300 to 550. To obtain these results, the simulations were initiated from a uniplanar-symmetric flow corresponding to the fully developed state at $\Rey^i = 450$. The corresponding results obtained from an initially axisymmetric state (open circular symbols) are shown as well. Based on these results, the following picture of the transition sequence emerges. For $\Rey^i < 326$, the flow remains axisymmetric. For $\Rey^i \in [326, 366]$ and $\Rey^i \in [370, 468]$, axisymmetry breaks down, but the system evolves toward a unique asymmetric branch depending on $\Rey^i$. Specifically, regardless of the amplitude of initial disturbance, the stable asymmetric branch is always biplanar within the first interval and uniplanar within the second. In the narrow range $\Rey^i \in (366, 370)$ and for $\Rey^i > 468$, both asymmetric branches coexist, and the selected branch depends on the initial disturbance amplitude. In particular, the biplanar branch is stable only to small disturbances, while the uniplanar branch is more likely to emerge when larger disturbances are present. A detailed discussion on the evolution of the lift coefficient in the first bistable regime, where $\Rey^i \in (366, 370)$, is provided in appendix\,\ref{app:bis-tab1}. Determining a quantitative threshold for the disturbance required to promote a transition between the two branches is not straightforward. Nevertheless, we believe that in the regime where $\Rey^i > 468$, this threshold increases with increasing $\Rey^i$. Indeed, in our restarted simulations, the uniplanar branch is no longer stable when $\Rey^i$ exceeds approximately 1500, whereas it remains stable for $\Rey^i$ up to 3000 if the simulations start from the fully developed uniplanar state at $\Rey^i = 1000$, where the initial disturbance is larger ($C_L = 0.16$ at $\Rey^i = 1000$, compared with $C_L = 0.088$ at $\Rey^i = 450$).

Figure \ref{fig:cl_e_bistable2}$(b)$ presents the evolution of the drag coefficient $C_D$, defined as $F_D = C_D \pi R^2 \rho^e u_{rel}^2 / 2$, in the considered $\Rey^i$ range. The drag remains virtually unchanged for $\Rey^i$ up to 326, where the flow remains axisymmetric. This behaviour is consistent with previous findings \citep{feng2001drag, edelmann2017numerical, 2024_Shi_drop}, which demonstrated that the drag coefficient $C_D$ depends only weakly on the internal Reynolds number when the flow is constrained to be axisymmetric. Thus, the horizontal dashed line in figure \ref{fig:cl_e_bistable2}$(b)$, representing $C_D$ at $\Rey^i = 325$, can be regarded as the reference drag coefficient for the corresponding axisymmetric configuration (hereafter referred to as $C_D^{axi}$) at higher $\Rey^i$ for $(\mu^\ast, \Rey^e) = (0.5, 200)$. Using this reference value, it becomes clear that the transition from the axisymmetric to the asymmetric branch (whether biplanar or uniplanar) is always accompanied by an increase in drag. The magnitude of this increase, denoted as $\Delta C_D = C_D - C_D^{axi}$, initially rises smoothly as $\Rey^i$ surpasses 326, where the bifurcation is supercritical. As $\Rey^i$ increases further, $\Delta C_D$ shows a jump for the three critical $\Rey^i$ values (366, 370, and 468), where the flow system transits between asymmetric branches. Specifically, at $\Rey^i = 366$ and 370, the switch from biplanar to uniplanar symmetry leads to an \emph{increase} in $\Delta C_D$ of approximately 0.02, whereas at $\Rey^i = 468$, the transition results in a \emph{decrease} of approximately 0.03. This distinction highlights the fundamental difference between the flow regime transition at the last critical $\Rey^i$ and those at the first two.

\section{Discussion} \label{sec:diss}
\subsection{Mechanism of the primary wake instability} \label{sec:mec_bif}
We have shown in \S\,\ref{sec:io_bif} and further corroborated in \S\,\ref{sec:axi} that the rotational symmetry of the base flow breaks down when the maximum internal surface vorticity $\omega_s^i$ exceeds a $\Rey^i$-dependent threshold $\omega_c^i$. Since the whole problem is governed by three dimensionless parameters, namely, the viscosity ratio $\mu^\ast$, the external Reynolds number $\Rey^e$, and the internal Reynolds number $\Rey^i$, the threshold $\omega_c^i$ may also be interpreted as, a critical internal Reynolds number when considering a series of cases with fixed $\mu^\ast$ and $\Rey^e$. This scenario has been explored in detail in \S\,\ref{sec:tra_seq}, where we set $(\mu^\ast, \Rey^e)=(0.5, 200)$ and determined the response of the flow as $\Rey^i$ (hence $\omega_s^i$) increased. The axisymmetric base flow was found to become unstable at $\Rey^i \approx 326$, beyond which it first transits to a steady, biplanar symmetric flow through a supercritical bifurcation. The biplanar flow structure post-bifurcation suggests that the steady mode with azimuthal wavenumber $m=2$ is responsible for breaking the axisymmetry. In the following, we elaborate in more detail on the mechanism by which, once produced into the base axisymmetric flow, the azimuthal vorticity may lead to its destabilisation.

As internal vorticity plays a key role in the onset of instability, it is relevant to examine its distribution within the droplet in the base flow, particularly near the instability threshold. As seen in figure \ref{fig:axi_streamline}$(b)$, for cases with fixed $\mu^\ast$ and $\Rey^e$, the isocontours of the azimuthal vorticity $\omega_\phi$ tilt towards the front stagnation point as $\Rey^i$ increases. By analysing additional results, we observed the same trend when increasing $\Rey^e$ (for fixed $\mu^\ast$ and $\Rey^i$) or decreasing $\mu^\ast$ (for fixed $\Rey^e$ and $\Rey^i$) (not shown). These findings, combined with the fact that the maximum internal vorticity $\omega_s^i$ increases with increasing $\Rey^e$ and $\Rey^i$ and decreasing $\mu^\ast$, indicate that a higher $\omega_s^i$ is associated with a more pronounced tilting of the $\omega_\phi$-isocontours inside the droplet.

\begin{figure}
\centerline{\includegraphics[scale=0.65]{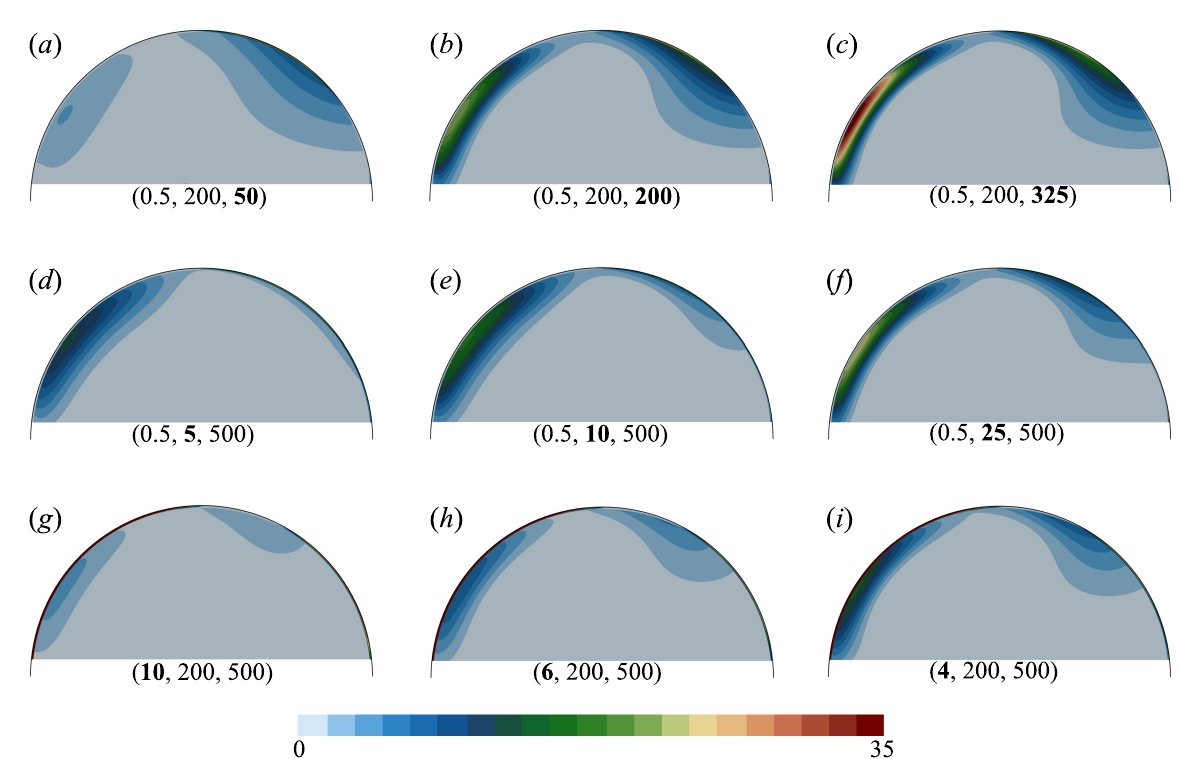}}
\caption{Isocontours of the normalised streamwise gradient of the azimuthal vorticity, $\p \omega_\phi/\p x \,(R^2/u_{rel})$, inside the droplet. The three numbers in brackets at the bottom of each panel correspond to $(\mu^\ast, \Rey^e, \Rey^i)$. Specifically, $\Rey^i$ increases from 50 to 325 from $(a)$ to $(c)$, $\Rey^e$ increases from 5 to 25 from $(d)$ to $(f)$, and $\mu^\ast$ decreases from 10 to 4 from $(g)$ to $(i)$. In each row, the last panel corresponds to the case closest to the onset of instability. In all panels, the ambient flow is directed from left to right.}
\label{fig:grad_vor}
\end{figure}
\begin{figure}
\centerline{\includegraphics[scale=0.595]{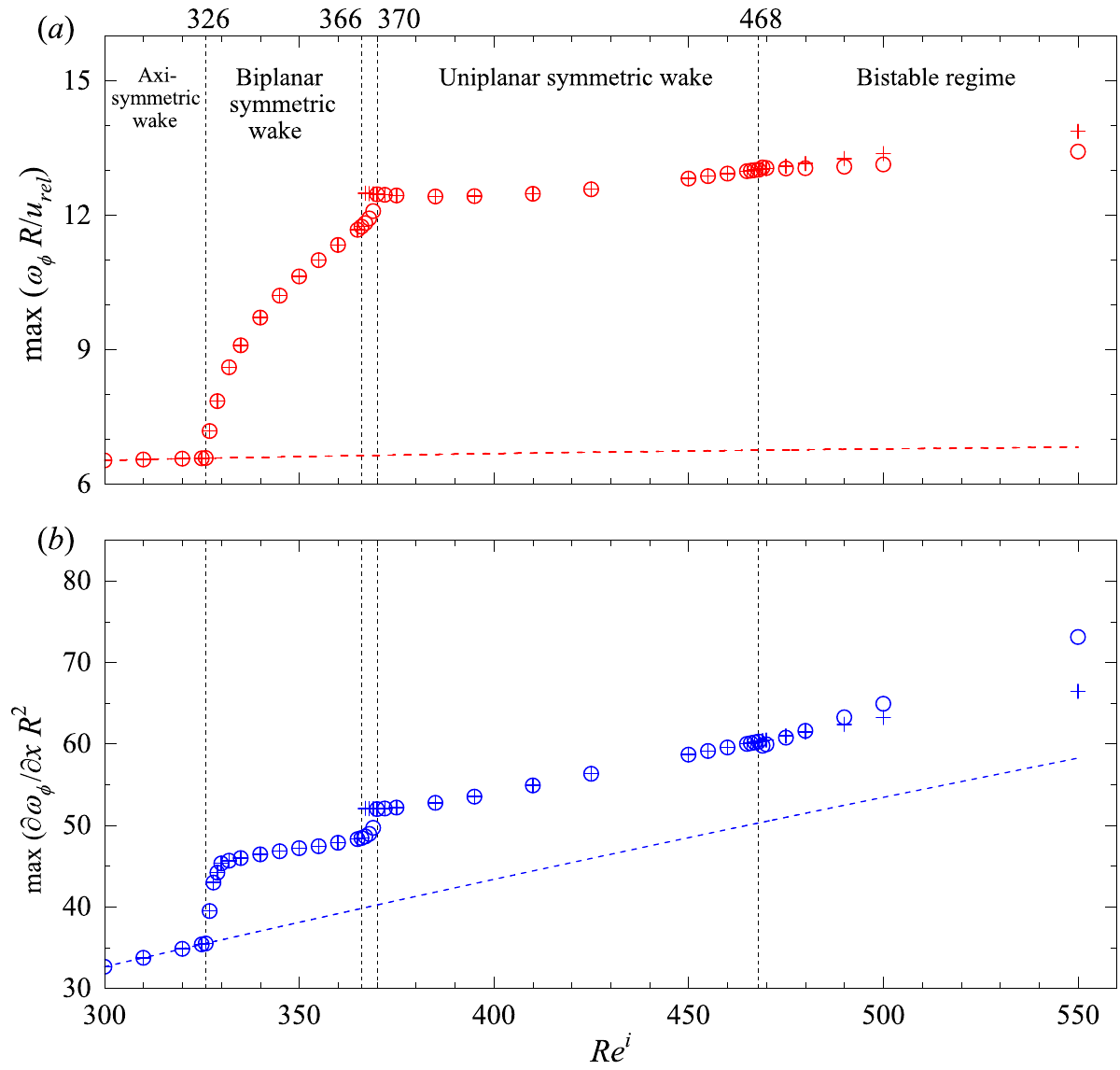}}
\caption{Maximum values of $(a)$ the normalised azimuthal vorticity, $\omega_\phi \,(R/u_{rel})$, and $(b)$ its streamwise gradient, $\p \omega_\phi/\p x \,(R^2/u_{rel})$, inside the droplet as a function of the internal Reynolds number for $(\mu^\ast, \,\Rey^e)=(0.5, \,200)$. In both panels, coloured dashed lines denote results from an axisymmetric flow configuration. $\bigcirc$: simulations starting from an initially axisymmetric flow; $+$: simulations starting from a uniplanar-symmetric flow corresponding to the fully developed state at $\Rey^i = 450$.}
\label{fig:omega_bistable}
\end{figure}
The tilting of the internal $\omega_\phi$-isocontours toward the front stagnation point as $\omega_s^i$ increases is an insightful observation, as a similar topological change has been reported for the external $\omega_\phi$-isocontours in the near wake of oblate spheroidal bubbles \citep{2007_Magnaudet, yang2007linear}. In that configuration, the external $\omega_\phi$-isocontours tend to align nearly perpendicular to the symmetry axis as $\Rey^e$ or the aspect ratio $\upchi$ (the ratio of the major to minor axes) increases. As this trend progresses, the streamwise gradient of $\omega_\phi$ must become increasingly pronounced for the viscous term $\nu^e \p^2\omega_\phi/\p x^2$ to counterbalance the inertial terms. Such a scenario is inherently unstable, leading to the breakdown of axisymmetry in the base flow when $\Rey^e$ exceeds a critical value for a given $\upchi$ (or vice versa).

In analogy to the argument above, the tilting of the internal $\omega_\phi$-isocontours observed in figure \ref{fig:axi_streamline}$(b)$ should correspond to a gradual increase in the streamwise gradient of $\omega_\phi$ as $\Rey^i$ increases. To confirm this, we present in figures \ref{fig:grad_vor}$(a-c)$ the isocontours of $\p \omega_\phi/\p x$ for increasing $\Rey^i$. Clearly, the maximum $\p \omega_\phi/\p x$ (normalised by $u_{rel}/R^2$) near the front of the droplet increases from 4.6 to 35.4 as $\Rey^i$ rises from 50 to 325. A similar trend is observed with increasing $\Rey^e$ [see figures \ref{fig:grad_vor}$(d-f)$, where $(\mu^\ast, \Rey^i) = (0.5, 500)$] and with decreasing $\mu^\ast$ [see figures \ref{fig:grad_vor}$(g-i)$, where $(\Rey^e, \Rey^i) = (200, 500)$]. These results confirm that, as $\omega_s^i$ increases, both the tilting of the $\omega_\phi$-isocontours and the streamwise gradient of the azimuthal vorticity within the droplet become more pronounced. Following the argument by \citet{2007_Magnaudet}, the internal flow can no longer remain stable if this gradient becomes sufficiently large. Indeed, although not explicitly shown, an inspection of the flow field at the onset of internal bifurcation reveals that the region where disturbances initially grow is closely aligned with the axial and meridional coordinates of the maximum $\p \omega_\phi/\p x$ in the corresponding axisymmetric configuration (see, for instance, the vortical threads near the front part of the droplet surface in the first inset of figure \ref{fig:bif_int}$b$). This serves as indirect evidence supporting the proposed relation between internal flow instability and the maximum streamwise gradient of the azimuthal vorticity.

While the relationship between internal flow instability and the presence of a sufficiently large $\p \omega_\phi/\p x$ proposed here is specific to axisymmetric flows (relevant to the first instability), we suspect that $\p \omega_\phi/\p x$ may still play a role in subsequent flow bifurcations, where the flow is already three-dimensional. Figure \ref{fig:omega_bistable} shows the evolution of the maxima of both $\omega_\phi$ and $\p \omega_\phi/\p x$ inside the droplet as a function of $\Rey^i$ for the series of cases with $(\mu^\ast,\,\Rey^e)=(0.5,\,200)$. Clearly, $\p \omega_\phi/\p x$ continues to increase significantly with $\Rey^i$ throughout the considered range, even when the flow is imposed as axisymmetric. In contrast, the maximum azimuthal vorticity remains nearly constant once the flow first transits from the biplanar to the uniplanar branch. These different behaviours are likely related to variations in $C_D$ as the flow transits at the two larger critical $\Rey^i$ values (see discussion in the last paragraph of the previous section) and highlight the consistent role of $\p \omega_\phi/\p x$ in triggering subsequent flow instabilities. Of course, once the flow becomes three-dimensional, both the streamwise and polar vorticity components become non-zero, and for a comprehensive understanding of flow instability, it may not be sufficient to examine only the azimuthal vorticity component. A rigorous stability analysis of the internal flow field, such as those conducted by \cite{yang2007linear} and \cite{tchoufag2013linear}, would be required to fully elucidate the mechanisms at play. Such an analysis would also help verify whether, at the first bifurcation, it is indeed the steady azimuthal mode with wavenumber $m=2$ that is amplified, leading to a non-axisymmetric but still steady flow. Conducting such an analysis, though beyond the scope of the present study, would be a valuable endeavour for future research.

\subsection{Relation between wake and path instabilities of a freely moving droplet of low-to-moderate $\mu^\ast$} 
\label{sec:mec_bif2}

In the previous sections, the droplet was considered to be spherical, with its centroid assumed to remain fixed. However, under physically realistic conditions, droplets can deform while moving freely under the effect of buoyancy and gravity. A key question is whether the internal flow bifurcation persists in this more general configuration and whether it plays a role in triggering the first path instability.

To provide insights into this question, we carry out three-dimensional simulations of a buoyant, deformable drop rising freely in an immiscible liquid that is otherwise at rest. The physical properties of the liquid-liquid system are selected to match those of a Toluene droplet rising in water, as investigated experimentally by \citet{wegener2010terminal}. The droplet radius is set to $R=1.2\,\text{mm}$, which is slightly larger than the threshold reported in the experiment. Unlike all simulations discussed so far, which were performed using the JADIM code, the simulations discussed below are carried out using the open-source flow solver \emph{Basilisk} \citep{popinet2009accurate, popinet2015quadtree}. This choice allows us to take advantage of the adaptive mesh refinement (AMR) technique built into the solver \citep{2018_Hooft} and to maintain a resolution at the interface comparable to that provided by the boundary-fitted mesh used earlier. Details of the numerical schemes used in this solver are provided in \citet{popinet2009accurate, popinet2015quadtree} and summarised in \citet{2024_Shi_bub}.

The one-fluid approach, combined with the geometric volume-of-fluid (VOF) method, is used to track and evolve the liquid-liquid interface. The computational domain is a cubic box with an edge length of $240R$. A free-slip condition is imposed on all four lateral boundaries, while a periodic condition is applied to the top and bottom boundaries \citep{2021_Zhang}. The spatial resolution is refined to approximately $1/68 R$ near the interface and to about $1/17R$ in the far wake (starting approximately $10R$ downstream of the droplet). The accuracy provided by the grid resolution is confirmed through a grid-independence study detailed in Appendix A of \citet{2025_Shi}, where the reliability and accuracy of the numerical method have been validated for freely rising objects at Reynolds numbers (both $\Rey^e$ and $\Rey^i$) up to approximately 1000. The droplet is initially spherical and is released midway between the four lateral boundaries and $15R$ above the bottom of the simulation domain.
\begin{figure}
\centerline{\includegraphics[scale=0.6]{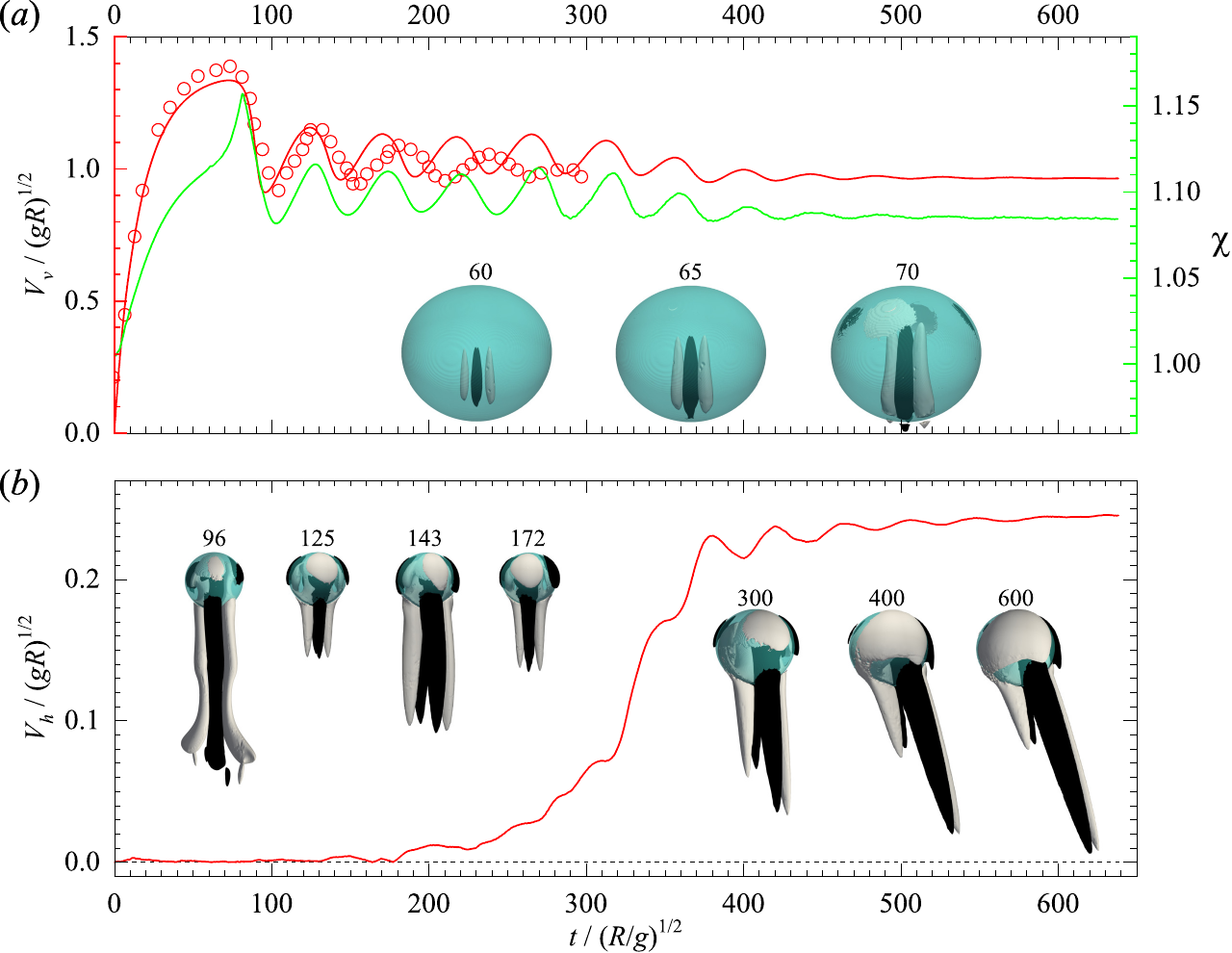}}
\caption{Time evolution of $(a)$ the vertical velocity ($V_v$, red line, left axis) and droplet aspect ratio ($\upchi$, green line, right axis), as well as $(b)$ the horizontal velocity ($V_h$) for a single Toluene droplet of radius $R=1.2\,\text{mm}$ rising in quiescent water [for detailed physical parameters, see Table 2 of \citet{wegener2010terminal}]. Both $V_v$ and $V_h$ are normalised by $(gR)^{1/2}$. In both panels, solid lines represent the present simulation results. In $(a)$, red open symbols denote experimental data of $V_v$ from \citet{wegener2010terminal} for $t/(R/g)^{1/2}$ up to 300, beyond which wall effects in the experiment significantly influenced the rising speed. The insets display the isosurfaces of the vertical component of the vorticity, $\omega_v (R/g)^{1/2}=\pm0.5$ (grey and black threads denote positive and negative values, respectively), at selected time instants (indicated at the top of each panel; values normalised by $(R/g)^{1/2}$). In all insets, the gravitational acceleration points vertically downwards, such that the droplet initially rises vertically upwards until approximately $t/(R/g)^{1/2} = 200$.}
\label{fig:2010_W}
\end{figure}

Figure \ref{fig:2010_W}$(a)$ compares the time evolution of the vertical velocity of the droplet, $V_v$, obtained from our three-dimensional simulation (solid line in red) with the corresponding experimental data (red symbols) from \citet{wegener2010terminal}. In both cases, the rising speed initially increases to a maximum at $t/(R/g)^{1/2} \approx 75$ before decreasing and oscillating around a mean value of approximately $1.0(gR)^{1/2}$ for $t/(R/g)^{1/2} \leq 250$. The first maximum of the rising speed and the reduced frequency (or Strouhal number, $St = 2fR/V_m$, where $V_m$ is the mean rising speed over the first two oscillation cycles and $f$ is the corresponding mean oscillation frequency) are $V_v^{\max} = 1.34(gR)^{1/2}$ and $St = 0.043$ in the simulation, closely consistent with the experimental values of $V_v^{\max} = 1.39(gR)^{1/2}$ and $St = 0.037$. The slight difference in the oscillation frequency is likely due to confinement effects in the experiment. There, the droplet initially rises along the axis of a cylindrical domain of radius approximately $21R$ \citep[see p. 89]{wegener2009einfluss}, compared to $120R$ in the simulation. As a result of this difference, the subsequent rising behaviour of the droplet in the experiment differs significantly from that observed in the simulation. In the experiment, the droplet begins to migrate laterally at $t/(R/g)^{1/2} \approx 225$ and collides with the wall at $t/(R/g)^{1/2} \approx 360$ \citep[see Fig. 5.4 therein]{wegener2009einfluss} (experimental data near the collision are omitted from figure \ref{fig:2010_W}$a$). As expected, the wall imposes a retarding effect on the rise speed, which causes a damping of velocity oscillations \citep{magnaudet2003drag, zeng2005wall, 2024_Shi_bubble_fix}. In contrast, the lateral migration in the simulation begins earlier, at $t/(R/g)^{1/2} \approx 200$, as seen in figure \ref{fig:2010_W}$(b)$. Beyond this point, the horizontal velocity, $V_h$, grows in time while the rising velocity undergoes damped oscillations. Both velocity components stabilise beyond $t/(R/g)^{1/2} \approx 600$, with the droplet rising steadily along an oblique path at an angle of approximately $\tan^{-1} (0.25/0.96) \approx 15^\circ$. Notably, in this terminal state, the external Reynolds number is approximately $\Rey^e = 256$, and the force balance among drag, lift, buoyancy, and gravity yields drag and lift coefficients of $C_D = 0.374$ and $C_L = 0.10$, respectively. These coefficients agree well with their counterparts in the corresponding spherical fixed-drop configuration, where we obtain $C_D = 0.372$ and $C_L = 0.093$.

From the fully resolved three-dimensional simulation results, it is also possible to examine the time evolution of the droplet shape. Assuming that the droplet remains almost oblate spheroidal during its rise, its shape can be characterised using the aspect ratio, $\upchi$, which represents the ratio of major and minor axes. The evolution of $\upchi(t)$ according to our numerical results is shown in figure \ref{fig:2010_W}$(a)$ (green line and right vertical axis). Throughout the evolution, the level of deformation remains marginal. Specifically, the maximum aspect ratio is approximately 1.16 and is reached shortly after the onset of the internal bifurcation. Thereafter, $\upchi$ oscillates around 1.1 before stabilising at 1.08. The shape oscillations observed in the transient stages largely result from variations in the rise speed. Indeed, given that the droplet deformation is small, its deformation level depends linearly on the Weber number and, consequently, primarily on the square of the rising speed. As a result, $\upchi$ reaches a local maximum shortly after $V_v$ reaches its peak, and the same correspondence holds for the local minima of $\upchi$ and $V_v$. Aside from these $V_v$-induced oscillations, no additional shape oscillation modes are observed, further confirming that the path instability must be attributed to flow instability rather than to the onset of a specific unstable deformation instability. 

We now elaborate, based on the numerical results from Basilisk, the relationship between the internal flow bifurcation and the evolution of the droplet motion. By examining the time evolution of the internal and external azimuthal energies (not shown), we found that the axisymmetry of the flow breaks down due to an internal flow bifurcation at $t/(R/g)^{1/2} \approx 60$, i.e. before the abrupt decrease in $V_v(t)$. This feature is highlighted by the insets in figure \ref{fig:2010_W}$(a)$, which illustrate the evolution of the streamwise vorticity during the interval $60 \leq t/(R/g)^{1/2} \leq 70$. Similar to what was observed in figure \ref{fig:bif_int}$(b)$, the disturbance associated with the asymmetry initially develops and grows only inside the droplet. Moreover, this disturbance is constrained within four vortex threads of equal intensity, maintaining a biplanar symmetric structure during the transition. The insets on the left side of figure \ref{fig:2010_W}$(b)$ show the vortical structure during the time oscillations of $V_v$. These results indicate that the velocity oscillations are closely related to the unsteady development of the wake. Specifically, $V_v$ reaches a local maximum at, for example, $t/(R/g)^{1/2} = 125$ and $172$, where the vortices shrink to their minimum extent, and the reverse occurs when $V_v$ reaches a local minimum, such as at $t/(R/g)^{1/2} = 96$ and $143$. The flow transits from biplanar symmetric to uniplanar symmetric at $t/(R/g)^{1/2} \approx 200$. As seen in the three rightmost insets of figure \ref{fig:2010_W}$(b)$, during this secondary transition, the left-right asymmetry grows over time, similar to the up-down asymmetry observed for a fixed droplet in figures \ref{fig:375i_vor_x} and \ref{fig:469i_vor_x}$(b)$. This asymmetry leads to a lift force that drives the lateral migration of the droplet.

\begin{figure}
\centerline{\includegraphics[scale=0.575]{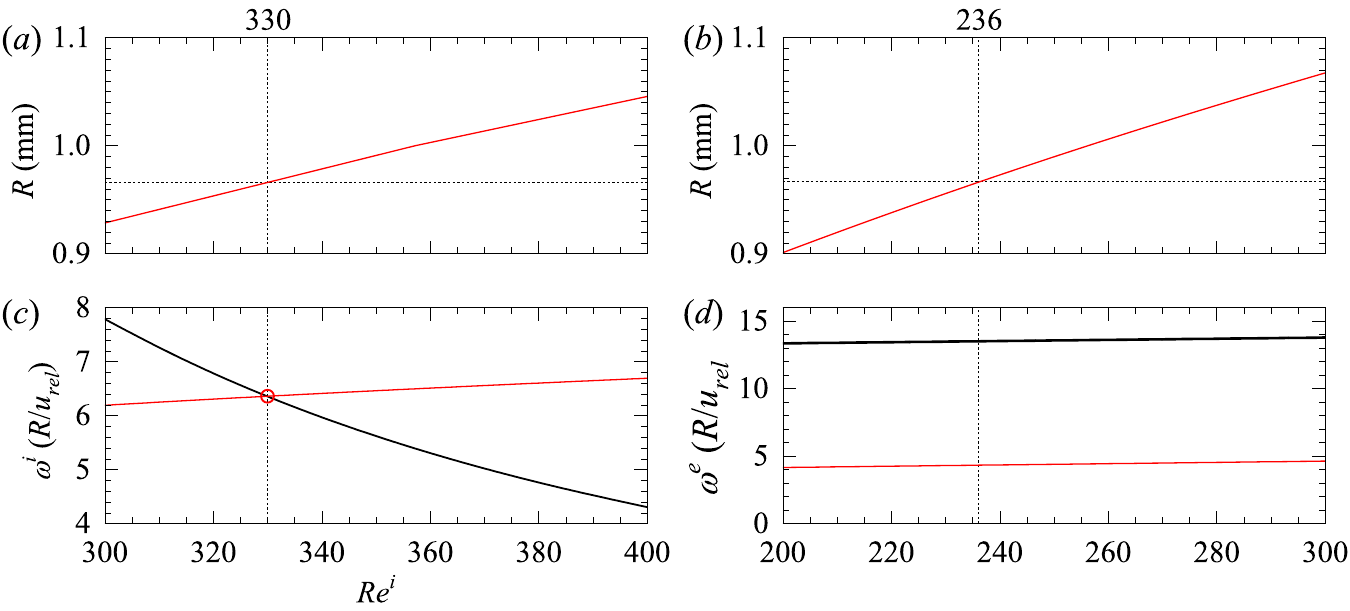}}
\caption{Characteristic parameters obtained in an axisymmetric configuration for Toluene droplets of sizes close to the threshold of the first path instability. $(a)$ (respectively $(b)$): Variation of the internal (respectively external) Reynolds number (horizontal axis) as a function of droplet radius $R$ (vertical axis). $(c)$ (respectively $(d)$): Maximum internal (respectively external) surface vorticity (red line) as a function of $\Rey^i$ (respectively $\Rey^e$). In $(a)$, the tick labels at the bottom match those in $(c)$, and the same correspondence holds between $(b)$ and $(d)$. In $(c)$, the black solid line represents the criterion for internal flow bifurcation (Eq.\,\eqref{eq:cri_vor}), while in $(d)$, the black solid line corresponds to the criterion for external flow bifurcation from \citet{2007_Magnaudet}.}
\label{fig:deform_drop}
\end{figure}

Based on the confirmed relationship between the internal flow bifurcation and the first path instability, we now assess the applicability of the empirical criterion proposed in \S\,\ref{sec:io_bif} (i.e. Eq.\,\eqref{eq:cri_vor}) for predicting the threshold droplet size at which path instability occurs first. To apply criterion \eqref{eq:cri_vor}, data on the maximum internal vorticity, $\omega_s^i$, and the internal Reynolds number, $\Rey^i$—both of which depend on the droplet radius, $R$—are required. To obtain these data, we conducted an additional series of simulations using Basilisk, considering the same liquid-liquid system while imposing axisymmetry on the flow field. The droplet radius was varied from $0.5$ to $1.5\,\text{mm}$ in increments of $0.05\,\text{mm}$. Figure \ref{fig:deform_drop}$(a)$ shows the variation of $\Rey^i$ (horizontal axis) as a function of the droplet radius (vertical axis) for $R$ near the path instability threshold. The corresponding results for $\omega_s^i$ are shown in panel $(c)$ (red line) as a function of $\Rey^i$. The intersection of the curve $\omega_s^i(\Rey^i)$ with the criterion \eqref{eq:cri_vor} (black solid line in panel $c$) corresponds to a critical internal Reynolds number of 330, which translates to a critical radius of $R=0.97\,\text{mm}$, as indicated in panel $(a)$. This closely matches the experimentally observed threshold of $R\approx1.1\,\text{mm}$ \citep{wegener2010terminal}. Furthermore, we recall that the critical radius for the onset of path instability was found to be $R\approx1.0\,\text{mm}$ in recent DNS by \citet{charin2019dynamic}. The numerical results for $\omega_s^e(R)$ and $\Rey^e(R)$, also obtained from these axisymmetric simulations, allow us to examine whether the external flow remains stable. Panel $(b)$ presents the external Reynolds number as a function of $R$ near the threshold of the internal flow bifurcation, with the critical radius corresponding to $\Rey^e=236$. Panel $(d)$ compares the maximum external surface vorticity (red line) with the threshold predicted by \citet{2007_Magnaudet} for external flow bifurcation (black solid line). At $\Rey^e=236$, the normalised maximum external surface vorticity, $\omega_s^e/(u_{rel}/R)$, is approximately $4.7$, which is only one-third of the threshold value ($13.5$). Hence, the external flow remains stable at the critical droplet size of path instability.

\subsection{Threshold droplet radius for internal bifurcation of a nearly spherical droplet moving in water}

\label{sec:reg_map}
Based on the key findings above, reference values for the critical droplet size required for the onset of internal flow bifurcation in a fluid-fluid system can be estimated. Specifically, for an immiscible liquid droplet freely moving in water, the critical radius, denoted as $R_c$, beyond which internal bifurcation may occur, can be determined by evaluating the maximum internal surface vorticity at steady state, where buoyancy, gravity, and drag forces are in equilibrium. 
\begin{figure}
\centerline{\includegraphics[scale=0.58]{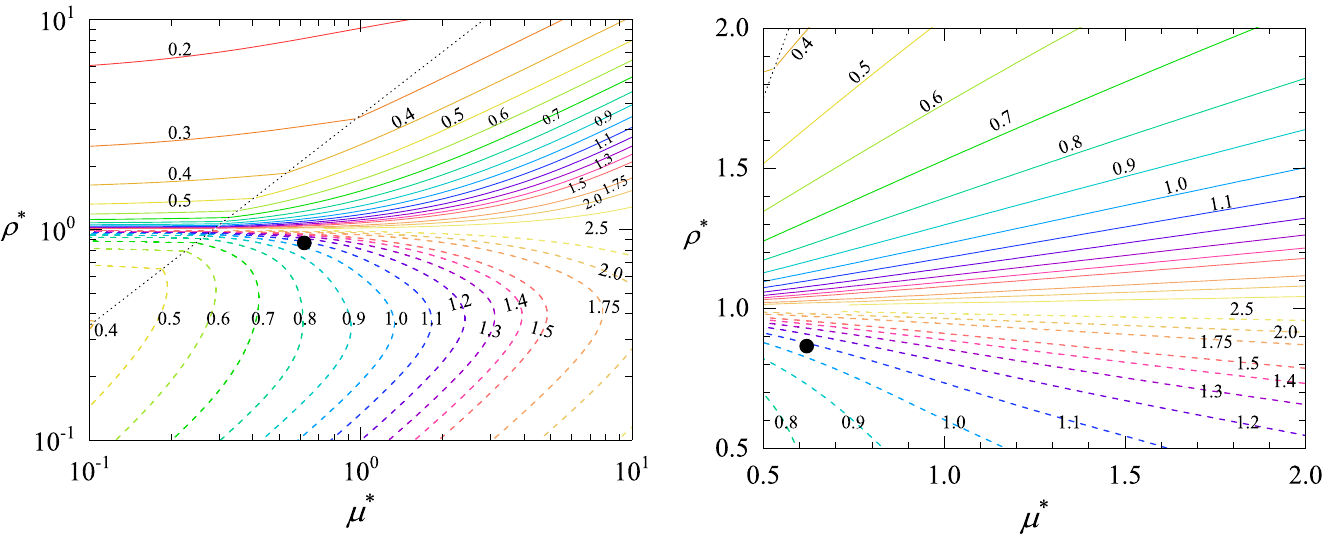}}
\caption{Threshold droplet radius, $R_c$ (in $\text{mm}$), in the $(\mu^\ast,\,\rho^\ast)$ phase plane for internal bifurcation in the case of a nearly spherical droplet rising (dashed lines) or settling (solid lines) in water. $(a)$: Results for $\mu^\ast$ and $\rho^\ast$ varying from 0.1 to 10. $(b)$: Same as $(a)$ but for $\mu^\ast, \rho^\ast \in [0.5,2]$. In both panels, coloured lines denote iso-$R_c$ contours, with values (in $\text{mm}$) indicated in the figure. The bullet symbol, located at $(\mu^\ast, \rho^\ast) = (0.62, 0.86)$, corresponds to the case of Toluene droplets in water, for which the critical $R_c$ is approximately $1.05\,\text{mm}$, as determined from the present regime map. In $(a)$, the dotted line in black corresponds to $\Rey^e = 100$ (or equivalently $\Rey^i = 350$), indicating that to the right (respectively, left) of this line, a sufficiently large $\Rey^i$ (respectively, $\Rey^e$) is required for internal flow bifurcation to occur. For details on the constraint of internal bifurcation in terms of $\Rey^e$ and $\Rey^i$, see \eqref{eq:force_1} in Appendix \ref{app:reg-map}.}
\label{fig:reg_map}
\end{figure}
A relatively coarse estimate of $R_c$ can also be obtained by considering the internal and external Reynolds numbers at steady state. Specifically, our fixed-droplet simulations indicate that internal bifurcation occurs typically for $\Rey^i \gtrsim 350$, provided that $\Rey^e = \mathcal{O}(100)$. Using these observations, we determined $R_c$ for various viscosity and density ratios (for details, see Appendix\,\ref{app:reg-map}), with both $\mu^\ast$ and $\rho^\ast$ ranging from 0.1 to 10, a parameter range commonly encountered for real liquid-liquid systems \citep{2020_Balla}. These results are summarised in figure \ref{fig:reg_map}$(a)$, which applies to droplets with small deformation, say, for $\upchi\leq1.1$. Together with the prerequisite $\Rey^e = \mathcal{O}(100)$, the range of validity may be further interpreted as an upper bound on the Morton number, $Mo$, up to approximately $3\times10^{-9}$, based on the empirical correlation by \citet[see Eq.\,(14) therein]{myint2007shapes}, where $Mo = g(\mu^e)^4|1-\rho^\ast|/(\rho^e\gamma^3)$ with $\gamma$ denoting the interfacial tension. Notably, within the considered range of $\mu^\ast$ and $\rho^\ast$, the critical $R_c$ varies from approximately $0.2\,\text{mm}$ to $2\,\text{mm}$, which is well within the typical size range of most practical systems \citep{2005_Clift}. Panel $(b)$ provides a zoomed-in view for $\mu^\ast, \rho^\ast \in [0.5,2]$. Within this refined parameter range, the minimum $R_c$ at a given $\mu^\ast$ is generally larger for a light droplet ($\rho^\ast < 1$) than for a heavy one ($\rho^\ast > 1$). For example, at $\mu^\ast = 1$, the minimum $R_c$ is approximately $0.95\,\text{mm}$ for a light droplet (attained at $\rho^\ast = 0.5$), while it is only about $0.5\,\text{mm}$ for a heavy droplet (attained at $\rho^\ast = 2.0$). Furthermore, for a Toluene droplet rising in water, a case with $(\mu^\ast, \rho^\ast) = (0.62, 0.86)$, panel $(b)$ indicates a critical radius $R_c$ of approximately $1.05\,\text{mm}$, which closely agrees with the experimental threshold reported by \citet{wegener2010terminal}.  

\section{Summary} \label{sec:summ}

We carried out three-dimensional numerical simulations of a uniform flow past a fixed spherical droplet over a wide range of governing parameters, namely, the viscosity ratio $\mu^\ast$, the external Reynolds number $\Rey^e$, and the internal Reynolds number $\Rey^i$. Our results show that for droplets with low-to-moderate viscosity ratios, the axisymmetric wake becomes unstable because of an internal flow bifurcation. This behaviour is absent for bubbles, particles, and droplets with large viscosity ratios, where the internal flow does not influence wake instability. The internal flow bifurcation originates from the surface vorticity produced at the external side of the droplet interface. By varying $\mu^\ast$, $\Rey^e$, and $\Rey^i$ independently, we found that the critical condition for the onset of internal bifurcation can be characterised in terms of the maximum vorticity on the internal side of the droplet surface, $\omega_s^i$. This leads to an empirical criterion based on a threshold $\omega_s^i$, denoted as $\omega_c^i$, which was found to depend solely on $\Rey^i$, to determine whether the axisymmetric flow remains stable.

Then, we selected a particular series of cases where $(\mu^\ast, \Rey^e)=(0.5, 200)$ to study the flow evolution with an increasing internal Reynolds number. Starting from an initially axisymmetric velocity field, the flow first undergoes a supercritical bifurcation at $\Rey^i=326$, yielding a steady non-axisymmetric flow that retains biplanar symmetry. In this configuration, the wake consists of two pairs of counter-rotating vortex threads of equal intensity, suggesting that symmetry breaking is associated with an azimuthal mode with wavenumber $m=2$ \citep{ghidersa2000breaking, yang2007linear}. With an additional increase in $\Rey^i$ beyond 370, a secondary bifurcation occurs, which is found to be subcritical. Following this transition, the flow loses its biplanar symmetry and becomes uniplanar symmetric, characterised by a single pair of counter-rotating vortices in the wake. Consequently, the droplet experiences a 
sizeable lift force, with $C_L$ showing an abrupt increase from a vanishingly small value to approximately 0.06 as $\Rey^i$ reaches 370. The secondary bifurcation persists up to $\Rey^i=468$, where $C_L$ further increases from 0.06 to about 0.09. For $\Rey^i > 468$, the flow reverts to a biplanar symmetric configuration. However, due to the subcritical nature of the secondary bifurcation, the final-state flow also depends on the initial disturbance amplitude. Specifically, by restarting all simulations from an asymmetric initial condition using the fully developed flow at $\Rey^i=450$ (where $C_L=0.088$), we identified two bistable regimes: a narrow range where $\Rey^i\in(366,370)$ and another for $\Rey^i>468$. In both bistable regimes, the final-state flow transits from biplanar to uniplanar symmetric when the initial disturbance exceeds a certain threshold.

Based on these findings, we proposed a physical explanation for the mechanism driving the primary wake instability. Examination of the azimuthal vorticity field in the base flow close to the threshold revealed that the isocontours inside the droplet tilt significantly toward the front, particularly when $\omega_s^i$ is large. This tilting is accompanied by a marked increase in the streamwise gradient of the internal azimuthal vorticity. Drawing an analogy with the argument proposed by \citet{2007_Magnaudet} for an external flow bifurcation, we suggested that if this streamwise gradient becomes sufficiently large, the internal flow cannot remain stable, leading to axisymmetry breakdown. Although this criterion probably provides only a sufficient condition for the primary wake instability, it aligns quantitatively with our numerical observations. A detailed stability analysis of the base flow in this regime is of course required to confirm the above scenario and obtain a more accurate criterion. 

Finally, we examine the relationship between the primary wake instability observed for a fixed droplet and the first path instability when the droplet is free to move. To this end, we conducted additional simulations of freely rising droplets, selecting physical parameters corresponding to those of a Toluene droplet rising in quiescent water \citep{wegener2010terminal}. Results from the three-dimensional simulations of a droplet with a radius of $R=1.2\,\text{mm}$ confirmed the onset of an internal flow bifurcation prior to the emergence of the first path instability, thereby establishing a direct connection between wake and path instabilities. Using the data for $(\omega_s^i, \Rey^i)$ obtained from constrained axisymmetric simulations over a wide range of droplet radii, along with the empirical criterion $\omega_c^i(\Rey^i)$ derived from the fixed-droplet simulations, we found that the predicted threshold droplet size for the primary wake instability closely matches the experimental and numerical thresholds for the onset of the first path instability \citep{wegener2010terminal, charin2019dynamic}. This further validates the proposed criterion for the internal flow bifurcation. Building on this confirmed relationship, we also estimated the threshold droplet size for the internal bifurcation of a nearly spherical droplet moving freely in water, using the criterion proposed in the present work.

One key aspect not addressed in this study is the mathematical nature of the secondary bifurcation that drives the transition from biplanar to uniplanar symmetry. Understanding this bifurcation is particularly important, as it leads to a lift force that, for a freely moving droplet, causes the transition from a vertical to an oblique path. Addressing this issue requires the development of a suitable global linear stability approach. Recent numerical techniques have made it possible to determine the threshold and nature of bifurcations in freely rising bubbles \citep{2024_Bonnefis}. Extending this approach to systems where the internal and external flow fields are coupled through kinematic and dynamic boundary conditions appears to be a promising next step to gain deeper insight into this fundamental problem.

\appendix
\section{The first bistable regime encountered with increasing $\Rey^i$ for $(\mu^\ast, \Rey^e)=(0.5, 200)$}\label{app:bis-tab1}
For the transition sequence discussed in \S\,\ref{sec:tra_seq}, a bistable regime exists within the narrow interval $\Rey^i\in(366,370)$, where the final-state flow structure can be either biplanar symmetric or uniplanar symmetric, depending on the amplitude of the initial disturbance. 

\begin{figure}
\centerline{\includegraphics[scale=0.575]{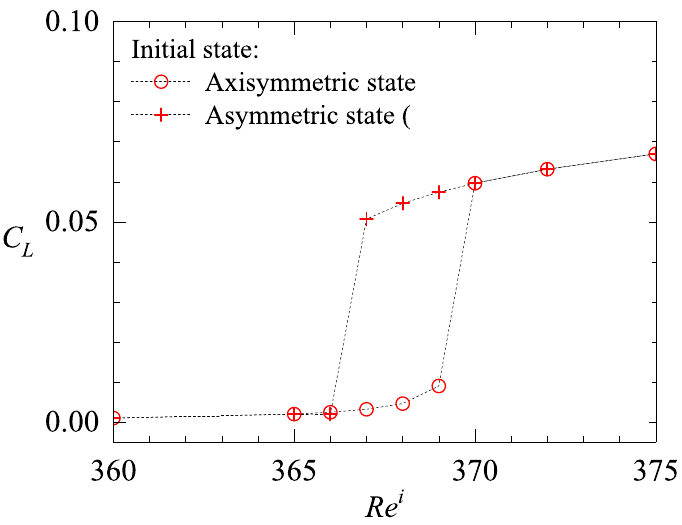}}
\caption{Lift coefficient obtained from different initial conditions for $\Rey^i$ increasing from 360 to 375. Circles ($\circ$) correspond to cases initialized from an axisymmetric flow, while plus signs ($+$) denote cases starting from an initially asymmetric flow based on the final-state result for $\Rey^i=450$ (where $C_L=0.088$).}
\label{fig:cl_bistab_1}
\end{figure}

\begin{figure}
\centerline{\includegraphics[scale=0.575]{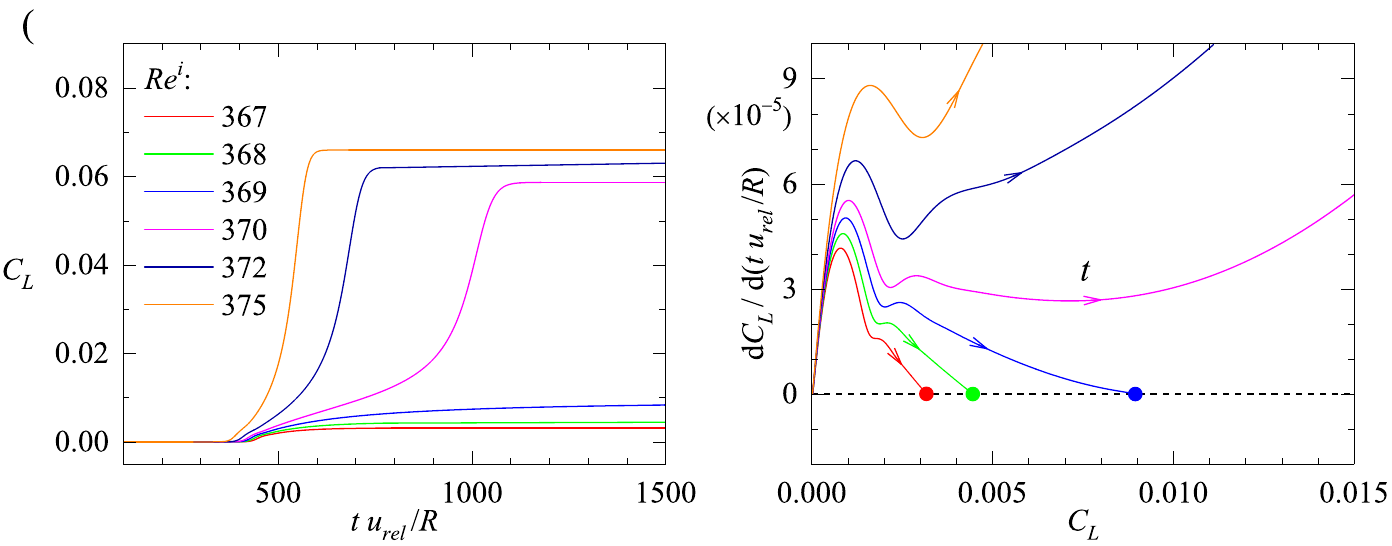}}
\caption{Same as figure \ref{fig:cl_rate_bistable} but for $\Rey^i$ increasing from 367 to 375. In $(a)$, the time evolution of the lift coefficient $C_L$ is shown for cases starting from an initially axisymmetric velocity field. In $(b)$, the variation of $\mathrm{d}C_L(t)/\mathrm{d}\left(t u_{rel}/R \right)$ as a function of $C_L(t)$ is presented, illustrating the emergence of a local fixed point near $C_L \approx 0.01$ for $\Rey^i = 369$.}
\label{fig:cl_rate_bistable2}
\end{figure}
Figure \ref{fig:cl_bistab_1} shows the lift coefficient obtained from different initial states for $\Rey^i$ increasing from 360 to 375. For cases starting from an axisymmetric flow, the final-state flow remains weakly biplanar symmetric up to $\Rey^i=369$, where $C_L$ is merely equal to 0.01. As $\Rey^i$ increases further, $C_L$ undergoes an abrupt rise to approximately 0.06 at $\Rey^i=370$, beyond which it shows a weak increase. Now, the above cases were carried out again but starting from an initially asymmetric velocity field, using the final-state result from $\Rey^i=450$ (where $C_L=0.088$), we find that the abrupt increase starts at $\Rey^i\approx 367$, slightly lower than in the previous scenario. However, for both $\Rey^i\leq366$ and $\Rey^i\geq370$, the resulting $C_L$ in the final state (and hence the corresponding flow structure) is independent of the initial conditions.

Figure \ref{fig:cl_rate_bistable2}$(a,b)$ show the time evolution of $C_L$ and its rate of change (as a function of $C_L$), respectively, obtained from the series of simulations starting from an axisymmetric flow. The evolution of $C_L(t)$ highlights a bottleneck effect as $\Rey^i$ decreases to 370, similar to that observed in \S\,\ref{sec:bistab_regime} for $\Rey^i$ increasing beyond 465 (see figure \ref{fig:cl_rate_bistable}). As $\Rey^i$ decreases slightly to 369, a local fixed point with small $C_L$ (approximately 0.01) emerges (panel $b$). According to panel $(b)$, this local fixed point shifts rapidly toward $C_L=0$ as $\Rey^i$ is decreased further. For $\Rey^i\leq366$, this local fixed point becomes globally stable, meaning that it cannot be eliminated even if the simulations are initialized from a highly asymmetric flow corresponding to the final stage of $\Rey^i=450$ (where $C_L=0.088$).

\section{Regime map of internal bifurcation of a nearly spherical freely rising or settling droplet in water}\label{app:reg-map}

The discussion in \S\,\ref{sec:mec_bif2} confirmed the close relationship between internal bifurcation and the first path instability of a freely rising droplet with a low-to-moderate viscosity ratio. Moreover, the criterion \eqref{eq:cri_vor}, based on the maximum internal surface vorticity, was found to predict reasonably well the threshold droplet size reported in experiments for a Toluene droplet rising in water. This agreement motivates us to construct a regime map for internal bifurcation in a general liquid-liquid system. To narrow the scope, we focus on the case of a freely rising or settling droplet in water under gravity. The key question we seek to address is: given the viscosity and density of the droplet (hence $\mu^\ast$ and $\rho^\ast$) \emph{a priori}, what is the threshold droplet radius corresponding to the onset of internal bifurcation?

To answer this question, one could first establish an empirical correlation to estimate the maximum internal surface vorticity, $\omega_s^i$, in the parameter space $(\mu^\ast, \Rey^e, \Rey^i)$. The corresponding external and internal Reynolds numbers, $(\Rey^e, \Rey^i)$, could then be determined for a droplet moving in water with radius $R$ with given viscosity and density ratios $(\mu^\ast, \rho^\ast)$. By equating $\omega_s^i$, now expressed as a function of $(R, \mu^\ast, \rho^\ast)$, with the critical vorticity threshold $\omega_c^i$—which itself depends on $(R, \mu^\ast, \rho^\ast)$ through $\Rey^e$ and $\Rey^i$—one would obtain the critical droplet radius, $R_c$. While this approach is essential for practical applications, it requires substantial effort to derive a reliable empirical correlation for $\omega_s^i$ over the three-parameter space $(\mu^\ast, \Rey^e, \Rey^i)$. Given this complexity, we opt instead for a simplified estimate of $R_c$ based on the fixed-droplet DNS results available from the present study.

We begin by examining the dependence of the critical internal surface vorticity, $\omega_c^i$, on the internal Reynolds number, $\Rey^i$. Inspection of Eq.\,\eqref{eq:cri_vor} reveals that the most rapid variation occurs for $\Rey^i \lesssim 300$. At $\Rey^i = 350$, $\omega_c^i$ decreases to approximately $5.5\,u_{rel}/R$, which is generally lower than the resulting $\omega_s^i$ for droplets with $\mu^\ast = O(0.1-1)$ moving at $\Rey^e = O(100)$. Thus, as a rule of thumb, internal bifurcation is likely to occur when $\Rey^i$ exceeds approximately 350. Given this, the next step is to determine the critical droplet radius, $R_c$, required for $\Rey^i$ to exceed this threshold. More specifically, since $\Rey^i = \Rey^e \rho^\ast/\mu^\ast$, the problem reduces to finding $R_c$ such that
\begin{equation}
\left\{ 
\begin{aligned}
&\Rey^e \rho^\ast/\mu^\ast \approx 350 & & \text{for} \quad \Rey^e \gtrsim 100, \\
& & &\text{or} \\
&\Rey^e \rho^\ast/\mu^\ast \gtrsim 350 & & \text{for} \quad \Rey^e \approx 100.
\end{aligned}
\right.
\label{eq:force_1}
\end{equation}

We now examine the relationship between $\Rey^e$ and the droplet radius, $R$. Assuming that the flow remains axisymmetric in the terminal state and that no shape oscillations occur, the balance between the drag, gravity, and buoyancy forces reads:
\begin{equation}
\rho^e |\rho^\ast - 1| \frac{4}{3} \pi R^3 g = \frac{1}{2} \pi R^2 \rho^e C_D\, u_{rel}^2.
\label{eq:force_2}
\end{equation}
Rearranging this expression, we obtain
\begin{equation}
R = \sqrt[3]{\frac{3}{32} \, \frac{\left( \mu^e \right)^2}{\left( \rho^e \right)^2 |\rho^\ast - 1| g} \, C_D \left( \Rey^e \right)^2}.
\label{eq:force_3}
\end{equation}
To employ Eq.\,\eqref{eq:force_3}, an appropriate correlation for the drag coefficient, $C_D$, is required, as it generally depends on $(\mu^\ast, \rho^\ast, \Rey^e)$. Assuming weak deformation, a reliable correlation for $C_D$ can be found in our recent work \citep{2024_Shi_drop}, where the hydrodynamic force on spherical droplets was examined. It was shown that, in the absence of internal bifurcation, $C_D$ depends only weakly on $\rho^\ast$ and can be approximated as:
\begin{equation}
 C_D  =  C_D^B + (R_\mu^m - 1) \frac{C_D^S - C_D^B}{(3/2)^m - 1},
\label{eq:force_4}
\end{equation}
where $R_\mu = (2 + 3\mu^\ast)/(2 + 2\mu^\ast)$ represents the intensity of the Stokeslet, and $C_D^B$ and $C_D^S$ denote the drag coefficients in the clean-bubble ($\mu^\ast \to 0$) and solid-sphere ($\mu^\ast \to \infty$) limits, respectively. The exponent $m$ is a fitted function of $\Rey^e$ and, along with $C_D^B$ and $C_D^S$, takes the following expressions:
\begin{subeqnarray} 
   C_D^B & = & \frac{16}{\Rey^e} \left\{  1 + \left[ \frac{8}{\Rey^e} + \frac{1}{2} \left(1 + 3.315\left(\Rey^e\right)^{-1/2}\right) \right]^{-1} \right\},\\[3pt]
  C_D^S & = & \frac{24}{\Rey^e} \left[ 1 + 0.15\left(\Rey^e\right)^{0.687} \right],\\[3pt]
  m & = & 1 + 0.01\left(\Rey^e\right)^{1.1}.
\label{eq:force_5}
\end{subeqnarray}

Now, substituting the expression for the critical $\Rey^e$ (given by Eq.\,\eqref{eq:force_1}) and the drag coefficient correlation (Eq.\,\eqref{eq:force_4} together with Eq.\,\eqref{eq:force_5}) into Eq.\,\eqref{eq:force_3}, and noting that the viscosity and density of water under standard conditions are approximately $\mu^e \approx 10^{-3} \, \text{Pa.s} $ and $\rho^e \approx 1000 \, \text{kg} \, \text{m}^{-3}$, respectively, we obtain solutions for $R_c$ over a broad range of viscosity and density ratios, $(\mu^\ast, \rho^\ast)$, with the results summarised in figure \ref{fig:reg_map}.


\backsection[Acknowledgements]{P. Shi gratefully acknowledges many fruitful discussions with J. Magnaudet, whose valuable insights have greatly contributed to understanding the possible mechanism of the primary wake instability. We also extend our gratitude to J. Zhang at Xi’an Jiaotong University for his generous support in implementing the periodic boundary condition in Basilisk. The computations were carried out on the HPC cluster \emph{hemera} at HZDR. }

\backsection[Funding]{This work is funded by the Deutsche Forschungsgemeinschaft (DFG, German Research Foundation) (P.S., grant number 501298479). }

\backsection[Declaration of interests]{The authors report no conflict of interest.}


\backsection[Author ORCIDs]{

Pengyu Shi, https://orcid.org/0000-0001-6402-4720; 

\'{E}ric Climent, https://orcid.org/0000-0001-9538-338X;

Dominique Legendre, https://orcid.org/0000-0002-6021-7119.
}


\bibliographystyle{jfm}
\bibliography{jfm}
\end{document}